\documentclass{jfm_modified_2016}
\usepackage{graphicx}
\usepackage{amsmath, amssymb}
\usepackage[inactive,tightpage]{preview}

\usepackage{microtype}
\usepackage[normalem]{ulem}
\usepackage{mathtools,bm}

\usepackage{esvect}

\usepackage{enumitem}

\pdfmapfile{+rsfso.map}
\DeclareSymbolFont{rsfso}{U}{rsfso}{m}{n}
\DeclareSymbolFontAlphabet{\mathscr}{rsfso}

\usepackage{pdflscape}
\usepackage{afterpage}
\usepackage{flafter}
\usepackage{rotating}
\usepackage{fancyhdr}
\fancypagestyle{lscape}{%
\fancyhf{} 
\fancyfoot[LE]{}
\fancyfoot[LO] {}
 
}

\usepackage{bm}
\usepackage{microtype}

\usepackage{hyperref}
\usepackage[usenames, dvipsnames]{xcolor}
\hypersetup{colorlinks=true,linkcolor=MidnightBlue,citecolor=MidnightBlue}
\usepackage{natbib}
\usepackage{array}
\usepackage{booktabs}
\usepackage{multirow}

\usepackage{caption}
\usepackage{subcaption}
\usepackage{tabularx}
\newcolumntype{Y}{>{\centering\arraybackslash}X}

\graphicspath{ {images/} }

\usepackage[percent]{overpic}
\usepackage[ruled,linesnumbered]{algorithm2e}
\usepackage{float}

\usepackage{tikz}

\newcommand*{\E}{\mathscr{E}}
\newcommand*{\Ehigh}{E_{\text{hw}}}
\renewcommand*{\i}{\mathrm{i}}
\newcommand*{\im}{\mathrm{i}}
\newcommand*{\e}{\mathrm{e}}

\renewcommand*{\Re}{\operatorname{Re}}
\renewcommand*{\Im}{\operatorname{Im}}
\newcommand*{\Arg}{\operatorname{Arg}}

\newcommand*{\de}{\operatorname{d\!}{}} 
\newcommand{\dd}[2]{\frac{\de#1}{\de#2}}

\usepackage{mathabx}
\usepackage{esint} 
\def\Xint#1{\mathchoice
   {\XXint\displaystyle\textstyle{#1}}%
   {\XXint\textstyle\scriptstyle{#1}}%
   {\XXint\scriptstyle\scriptscriptstyle{#1}}%
   {\XXint\scriptscriptstyle\scriptscriptstyle{#1}}%
   \!\int}
\def\XXint#1#2#3{{\setbox0=\hbox{$#1{#2#3}{\int}$}
     \vcenter{\hbox{$#2#3$}}\kern-.5\wd0}}

\def\YYint#1#2#3{{\setbox0=\hbox{$#1{#2#3}{\int}$}
     \vcenter{\hbox{\scalebox{1}[-1]{$#2#3$}}}\kern-.5\wd0}}

\def\dashint{\Xint-}

\newcommand*{\partamp}{2.2}

\shorttitle{Exponential asymptotics and parasitic capillary ripples}
\shortauthor{Shelton and Trinh}

\title{Exponential asymptotics for steady parasitic capillary ripples on steep gravity waves}

\author{Josh Shelton\corresp{\email{j.shelton@bath.ac.uk}}
 \and Philippe H. Trinh\corresp{\email{p.trinh@bath.ac.uk}}}

\affiliation{
Department of Mathematical Sciences, University of Bath, Bath BA2 7AY, UK
}

\pubyear{XXXX}
\volume{YYY}
\pagerange{xxx--xxx}
\date{\today~[Draft]}

\begin{document}
\maketitle

\begin{abstract}
In this paper we develop an asymptotic theory for steadily travelling gravity-capillary waves under the small-surface tension limit. In an accompanying work [\citealt{Shelton2021onthe}, \emph{J. Fluid Mech.}, vol. 922] it was demonstrated that solutions associated with a perturbation about a leading-order gravity wave (a Stokes wave) contain surface-tension-driven parasitic ripples with an exponentially-small amplitude. Thus a naive Poincar\'{e} expansion is insufficient for their description. Here, we shall develop specialised methodologies in exponential asymptotics for derivation of the parasitic ripples on periodic domains. The ripples are shown to arise in conjunction with Stokes lines and the Stokes phenomenon. The resultant analysis associates the production of parasitic ripples to the complex-valued singularities associated with the crest of a steep Stokes wave. A solvability condition is derived, showing that solutions of this type do not exist at certain values of the Bond number. The asymptotic results are compared to full numerical solutions and show excellent agreement. The work provides corrections and insight of a seminal theory on parasitic capillary waves first proposed by Longuet-Higgins [\textit{J. Fluid Mech.}, vol. 16 (1), 1963, pp. 138-159].
\end{abstract}

\section{Introduction}

\begin{figure}
\includegraphics[scale=15]{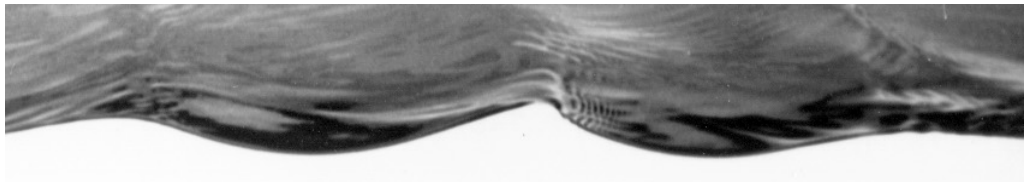}
\caption{\label{fig:ripples} Experimental picture showing parasitic ripples located near the crests of a steep gravity-dominated wave. Note that the ripples appear in an asymmetric manner; mechanisms that produce asymmetry are discussed in \S\ref{sec:asymmetry}. Image used with permission from Professor N. Ebuchi (Hokkaido University)}
\end{figure}

\noindent Consider the situation of a steep gravity-driven \textit{Stokes} wave---a two-dimensional periodic surface wave of an inviscid and irrotational fluid travelling without change of shape or form. If a small amount of surface tension is included, it is reasonable to expect that, under certain conditions, the profile of the Stokes wave is modified or perturbed by a small amount. Physically, such perturbations may manifest as small-amplitude capillary-driven ripples concentrated near the crest of the wave. We shall refer to these perturbations as \textit{parasitic ripples}, an experimental observation of which appears in figure~\ref{fig:ripples}.

The purpose of this work is to develop a precise asymptotic theory for the parasitic ripples that arise in the permanently progressive framework of a travelling water-wave. In particular, we shall demonstrate that for small surface tension, the parasitic ripples are described by an exponentially-small remainder to the base water-wave, which is given by a typical asymptotic expansion in algebraic powers of the surface tension parameter. Their description requires the use of exponential asymptotics, and indeed, it is this requirement that distinguishes this work from the previous analytical treatments.

\subsection{Steady parasitic solutions for small surface tension}
\label{sec:JS}

\noindent Here, we shall provide a brief overview of how our treatment differs from previous works. To begin, the water-wave problem can be formulated in terms of an unknown streamline speed, $q$, and streamline angle, $\theta$, considered as functions of the velocity potential, $\phi$, over the periodic domain $-\tfrac{1}{2}<\phi\leq \tfrac{1}{2}$. The free-surface is then governed by Bernoulli's equation,
\begin{equation}
\label{eq:Inequation}
F^2 q^2 \frac{\de q}{\de \phi}+ \sin{(\theta)}-Bq\frac{\de}{\de \phi}\bigg(q\frac{\de \theta}{\de \phi}\bigg)=0,
\end{equation}
where $F$ is the Froude number, and $B$ is the (inverse)-Bond number. These non-dimensional constants are given by
\begin{equation}
\label{eq:InFB}
F=\frac{c}{\sqrt{g \lambda}} \qquad \text{and} \qquad B = \frac{\sigma}{\rho g \lambda^2},
\end{equation}
where $c$ is the wave speed, $g$ is the constant acceleration due to gravity, $\lambda$ is the wavelength, $\rho$ is the fluid density, and $\sigma$ is the coefficient of surface tension. The limit of small-surface tension is given by $B \to 0$.

 \begin{table}
  \centering
  \scriptsize
  \begin{tabular}{c c p{9.5cm}}
  & Symbol   & Notes     \\[5pt]

 Dimensional & $c$ & Wave speed  \\
 quantities & $g$ & Constant acceleration due to gravity  \\
  & $\rho$ & Fluid density \\
  & $\lambda$ & Wavelength  \\
  & $\sigma$ & Constant coefficient of surface tension  \\[5pt]
  
  Parameters & $q$ & Streamline speed \\
  & $\theta$ & Streamline angle \\
  & $ \phi+ \i \psi$ & Complex potential comprised of velocity potential $\phi$ and streamfunction $\psi$  \\
  & $f$ & Complex valued domain, relabeled from the analytically continued velocity potential $\phi_c$ \\
  & $a$ & Direction of analytic continuation, where $a=\pm1 $ \\
  & $\E$ & Energy \\
  & $B$ & Bond number \\
  & $F$ & Froude number \\[5pt]

  Subscript & $x_{\phi}$ & Partial derivative of $x$ with respect to $\phi$  \\
  notation & $q_n$ & $n$th order of the asymptotic series $\sum_{n=0}^{\infty} B^n q_n$  \\
  & $E_{\text{hw}}$ & Text, used for hw (highest wave), homog. (homogeneous), and phys. (physical)  \\
  & $Q_a$ & Direction of analytic continuation of the free-surface solution, $Q(f)$ \\
[5pt]

  Further  & $\widehat{\mathscr{H}}$ & Complex-valued Hilbert transform  \\
  notation & $f^*$ & Location of the principal singularity of the analytically continued Stokes wave.  \\
  & $\bar{q}$ & Overbar, denoting the remainder to a truncated asymptotic series  \\
  & $\mathfrak{q}$ & Frankerscript, denoting the combined solution $q \rvert_{a=-1} + q \rvert_{a=1}$ \\
  & $\xi$ & Forcing terms which appear in the equation for the remainder, $\bar{q}$  \\
  & $\hat{q}$ & Hats denote an inner asymptotic solution within a boundary layer associated with the singularity at $f=af^*$   \\[5pt]

  \end{tabular}
  \caption{List of variables, parameters, and notation used in the main text.}
  \label{table:flex-bigtable}
  \end{table}

\begin{figure}
\includegraphics[scale=0.95]{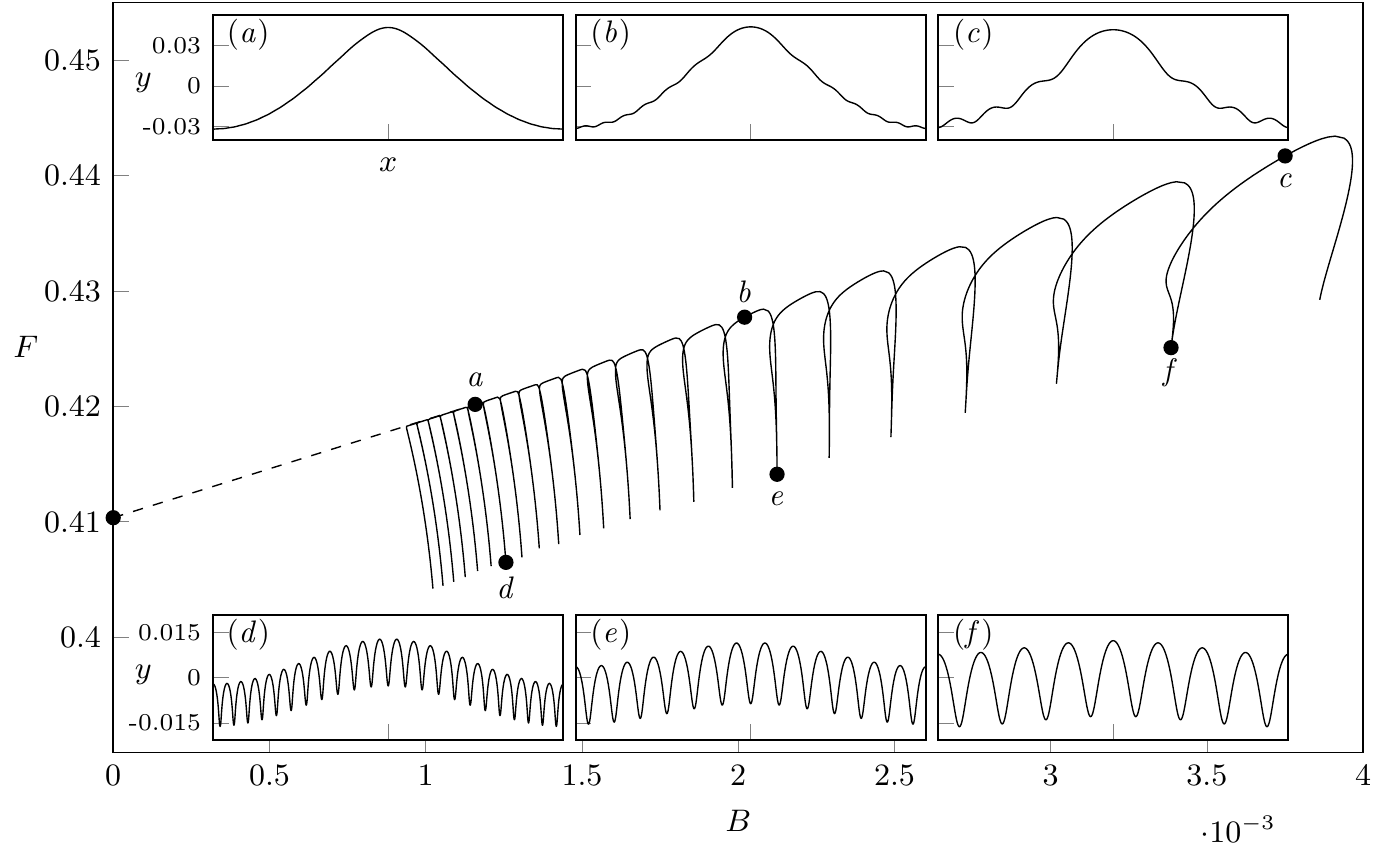}
\caption{\label{fig:JSbifurc} The numerical ($B,F$) solution space calculated by \cite{Shelton2021onthe} is shown for a fixed energy of $\E=0.3804$. The insets ($a$, $b$, $c$) show the physical free-surface for those cases corresponding to exponentially-small parasitic ripples on Stokes waves; the insets ($d$, $e$, $f$) show a different multiple-scales regime.}
\end{figure}

As it turns out, the structure of the solution space for the free-surface gravity-capillary wave problem is remarkably sophisticated. 
Recently, a portion of this solution space was investigated numerically by \cite{Shelton2021onthe} for fixed energy, with a focus on determining the small-surface tension limit of $B \to 0$.
Multiple branches of solutions were found, each of which can be indexed by the number of capillary-driven ripples that appear in the periodic domain. This solution space is shown in figure~\ref{fig:JSbifurc} and the structure of `fingers' (as introduced in the previous work) can be observed.

Two different asymptotic limits are visible in these solutions.
The first limit is observed from solutions $(d)$, $(e)$, and $(f)$ at the lower parts of each of the fingers, which are highly oscillatory with some modulation across the domain. In this region, the solution can be approximated by a multiple-scales framework, with 
\begin{equation}
 q(\phi)=\sum_{n=0}^{\infty} B^n q_n(\phi, \hat{\phi}),
\end{equation}
where $\hat{\phi}=\frac{\phi}{B}$ is the fast scale. Substitution of this ansatz into Bernoulli's equation \eqref{eq:Inequation} yields, at order $1/B$, the pure-capillary equation of \cite{crapper1957exact} for the small-scale ripples
\begin{equation}
\label{eq:CapEq}
F_0^2 q_0^2 \frac{\partial q_0}{\partial \hat{\phi}}- q_0 \frac{\partial}{\partial \hat{\phi}} \bigg( q_0 \frac{\partial \theta_0}{\partial \hat{\phi}}\bigg)=0.
\end{equation}
Thus for these multiple-scale solutions the highly oscillatory parasitic ripples appear in the leading order term, $q_0(\phi, \hat{\phi})$, of the expansion. We will focus upon this asymptotic regime in future work.

\begin{figure}
\centering
\includegraphics[scale=1]{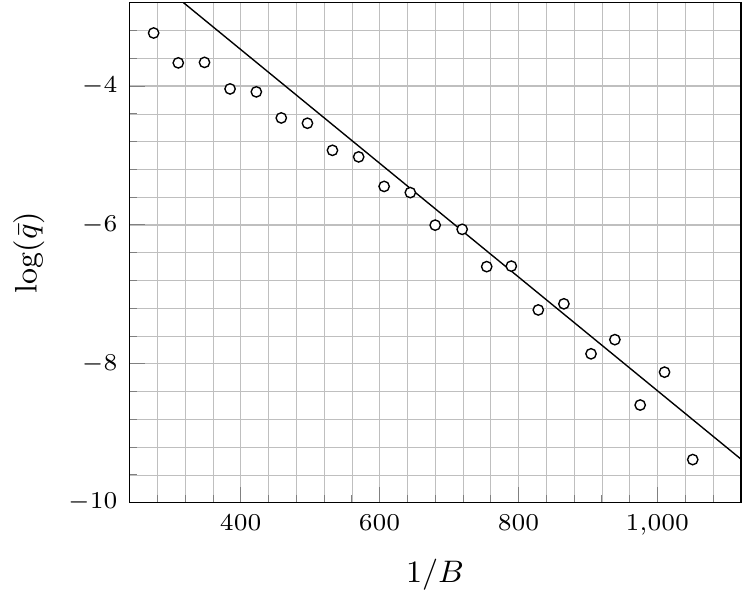}
\caption{\label{fig:expscale} Our analytical prediction of the exponential-scaling of the parasitic ripple magnitude, $\bar{q}$, (line) is compared with numerical results of the full nonlinear equations (circles). These results have both been calculated with an energy of $\E=0.3804$, and the gradient of the analytical result is $-0.0082$.}
\end{figure}

The second asymptotic limit can be observed in subfigures $(a)$, $(b)$, and $(c)$ of figure~\ref{fig:JSbifurc}. As these solutions approach the pure-gravity (Stokes) solution with the same fixed value of the energy as $B \to 0$, the leading order solution $q_0$ contains no ripples.
Moreover, a standard perturbative series of the form
\begin{equation}
 q(\phi)=\sum_{n=0}^{\infty} B^n q_n(\phi)
\end{equation}
will also not contain the parasitic-ripples observed in the numerical solutions. This is due to the exponential-smallness of the amplitude of these ripples, which was confirmed numerically by \cite{Shelton2021onthe} and is shown to form a straight line in the semi-log plot in figure~\ref{fig:expscale}.

Thus, in the $B \to 0$ limit, the capillary-driven ripples exhibit different behaviours according to two distinct asymptotic limits of: 
\begin{enumerate}[label=(\roman*),leftmargin=*, align = left, labelsep=\parindent, topsep=3pt, itemsep=2pt,itemindent=0pt ]
\item a multiple-scales solution, for which the ripples appear in the leading-order approximation of the solution; and 
\item a standard perturbative series about a Stokes wave, for which the parasitic ripples appear beyond-all-orders.
\end{enumerate}
It is this latter asymptotic regime that we will focus upon in this work.

In the context of the above second scenario, an early analytical theory for the generation of these parasitic ripples was proposed by \cite{longuet-higgins_1963_the_generation}, who considered a small surface-tension perturbation about a base Stokes wave. Although Longuet-Higgins' seminal work provides a crucial basis for our analysis in this paper, we shall also demonstrate that there are a number of key asymptotic inconsistencies that appear in the historical 1963 work. These inconsistencies turn out to be connected with modern understanding of exponential asymptotics (\citealt{berry1989uniform, olde1995stokes, chapman_1998_exponential_asymptotics}), and may have led to the poor agreement noted by \cite{perlin1993parasitic} in comparison with numerical solutions of the full nonlinear problem. One of the primary objectives of our work is to provide a critical re-examination of the seminal \cite{longuet-higgins_1963_the_generation} paper, which we perform in \S\ref{sec:LH}. Note that we shall provide a more complete literature review of theories and research on the parasitic capillary problem in our discussion of \S{\ref{sec:Di}}. 

As we shall demonstrate, the intricate difficulties involved in formulating a corrected theory for the $B \to 0$ limit are linked to the presence of singularities in the analytical continuation of the leading-order gravity-wave solution. Due to the singularly-perturbed nature of Bernoulli's equation \eqref{eq:Inequation}, successive terms in the asymptotic expansion of the solution require repeated differentiation of the singularity in the leading-order solution. This causes the expansion to diverge. In studying this divergence, a form for the exponentially small correction terms to the asymptotic series is found by truncating the series optimally and these corrections correspond to the anticipated parasitic ripples.

\subsection{Outline of the paper}

\noindent We begin in \S\ref{sec:formulation} with the mathematical formulation of the non-dimensional gravity-capillary wave system, which is analytically continued into the complex potential plane. 
In \S\ref{sec:LH} we provide a detailed overview of the  \cite{longuet-higgins_1963_the_generation} analytical methodology.
In \S\ref{sec:EA}, we consider a perturbation expansion for small values of the surface tension, $B$. Subsequent terms in this expansion rely on differentiation of the leading order gravity-wave solution. Thus, singularities in the analytic continuation of the free-surface gravity-wave produce a divergence in the asymptotic series as further terms are considered. The scaling of the principal upper-half and lower-half singularities are derived in \S\ref{sec:ACstokes}.
The divergence of the late-terms of the asymptotic expansion is then considered in \S\ref{sec:lateorders}. This allows us to find the Stokes lines for our problem, which are shown in \S\ref{sec:expasymp} to produce the switching of exponentially-small terms of the solution via Stokes phenomenon. Application of the periodicity conditions then yields an analytical solution for these parasitic ripples and an accompanying solvability condition.
These solutions and the solvability condition are then compared to numerical solutions of the full nonlinear equations in \S\ref{sec:Comparisons}.
Our findings are summarised in \S \ref{sec:Co}, and discussion of further work occurs in \S \ref{sec:Di}.

\section{Mathematical formulation}\label{sec:formulation}

\begin{figure}
\includegraphics[scale=1]{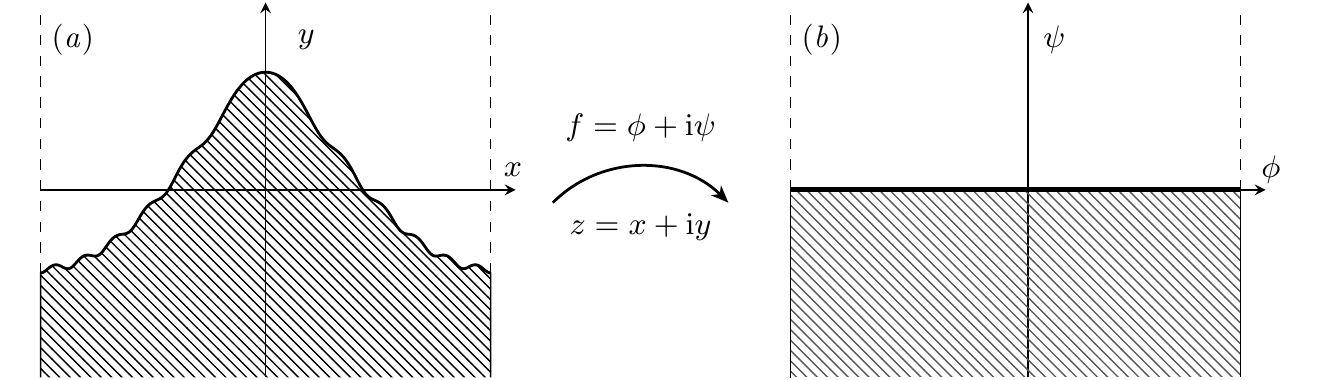}
\caption{\label{fig:phipsi} The conformal map from $(a)$, the physical $z = x + \im y$-plane, to $(b)$, the complex $f = \phi + \im\psi$-plane, is shown. The boundary, $y=\eta(x)$, is mapped to the line $\psi=0$.}
\end{figure}

\noindent We begin by considering the two-dimensional free-surface flow of an inviscid, irrotational, and incompressible fluid of infinite depth. The effects of gravity and surface tension are included. We assume the free-surface to be periodic with wavelength $\lambda$, and it is chosen to move to the right with wave speed $c$. Imposing a sub-flow within the fluid in the opposite direction cancels out the lateral movement; this results in a steady free-surface when $\partial_t=0$, now assumed to be located at $y=\eta(x)$. A typical configuration is shown in figure~\ref{fig:phipsi}. 
The system is non-dimensionalised using $\lambda$ and $c$ for the units of length and velocity, respectively, and the set of governing equations is taken to be the same as those considered by \cite{Shelton2021onthe}:
\begin{subequations} \label{eq:MF1}
\begin{align}
\phi_{xx}+\phi_{yy}=0& \qquad \text{for $y \leq \eta$}, \label{eq:laplace} \\
\phi_{y}=\eta_{x}\phi_{x}& \qquad \text{at  $~y=\eta$,} \label{eq:kin}\\
\frac{F^2}{2}(\phi^{2}_{x}+\phi^{2}_{y})+y-B\frac{\eta_{xx}}{(1+\eta_x^2)^{\frac{3}{2}}}=\frac{F^2}{2}& \qquad \text{at  $~y=\eta$,} \label{eq:dyn}\\
\phi_{y} \to 0 \quad \text{and} \quad \phi_{x} \to -1 & \qquad \text{as  $~y \to -\infty$}. \label{eq:deep}
\end{align}
Thus the flow is governed by Laplace's equation \eqref{eq:laplace}, kinematic and dynamic boundary conditions in \eqref{eq:kin} and \eqref{eq:dyn} respectively at the free-surface, and the deep-water condition \eqref{eq:deep}. The constants $F$ and $B$ are the Froude and Bond numbers, introduced earlier in equation \eqref{eq:InFB}. Periodicity of the flow and wave profile is specified by enforcing 
\begin{equation}\label{eq:periodic_new}
  \nabla \phi \left(x -\tfrac{1}{2}, y\right) = \nabla \phi\left(x + \tfrac{1}{2}, y\right) \quad \text{and} \quad 
  \eta\left(x - \tfrac{1}{2}\right) = \eta\left(x+\tfrac{1}{2}\right).
\end{equation}
\end{subequations}

In addition to the governing equations in \eqref{eq:MF1}, we also enforce an amplitude parameter as a measure of nonlinearity of the solution. This is derived from the physical bulk energy of the wave via Appendix~A of \cite{Shelton2021onthe}. This yields
\begin{equation} \label{eq:Energy}
\E=\frac{1}{\Ehigh}\int_{-\frac{1}{2}}^{\frac{1}{2}}\bigg[ \frac{F^2}{2}y(x_{\phi}-1)  +B \Big(\sqrt{(x_{\phi}^2+y_{\phi}^2)}-x_{\phi}\Big)+\frac{1}{2}y^2 x_{\phi} \bigg]  \de \phi,
\end{equation}
where the three groupings of terms correspond to the kinetic, capillary, and gravitational potential energies. In \eqref{eq:Energy} we have rescaled with the energy of the limiting classical Stokes wave, $\Ehigh \approx 0.00184$. A central idea in \cite{Shelton2021onthe} concerned the importance of choosing an amplitude condition on the water waves, and we refer readers to \S{\partamp} of that work for further discussion. 

Finally, based on the previous study in \cite{Shelton2021onthe}, we note that once the energy condition \eqref{eq:Energy} is imposed, there is only a single degree of freedom in specifying either $F$ or $B$. We typically consider the Bond number as a free parameter, which results in the Froude number as an eigenvalue that must be determined via the system \eqref{eq:MF1}.

\subsection{The $(q,\theta)$ formulation}\label{sec:qtheta}

\noindent In this section, we repose the two-dimensional governing system \eqref{eq:MF1} as a one-dimensional boundary-integral formulation in terms of the free-surface speed and angle.
Following the traditional treatment of potential free-surface flows, we introduce the complex potential $f=\phi+\i \psi$. Rather than consider $f = f(z)$, we instead consider $z = z(f)$, and hence the flow region is known in the potential plane. The complex potential plane is shown in figure~\ref{fig:phipsi}. From this definition, the complex velocity can be found to be $\de{f}/\de{z} = u-\i v$, where $(u,v)$ are the horizontal and vertical velocities.

Introducing $q$ as the streamline speed and $\theta$ as the streamline angle by the relationship $q \e^{-\i \theta}=u-\i v$ then yields
\begin{equation}\label{eq:cv}
\frac{\de f}{\de z}=q \e^{-{\mathrm{i}} \theta}.
\end{equation}
In this form, Bernoulli's equation \eqref{eq:dyn} is written as
\begin{subequations}
\begin{equation}\label{eq:Bern-qtheta}
F^2 q^2 \frac{\de q}{\de \phi}+ \sin{(\theta)}-Bq\frac{\de}{\de \phi}\bigg(q\frac{\de \theta}{\de \phi}\bigg)=0.
\end{equation}
By the analyticity of $\log{q}-\i \theta$, we introduce the boundary-integral equation which relates $q$ to the Hilbert transform of $\theta$ operating over the free-surface. For our periodic domain from $-1/2$ to $1/2$, we integrate $\log{q} - \im\theta$ using Cauchy's theorem and use the periodicity conditions
\begin{equation}\label{eq:periodicqth}
q \left(\phi -\tfrac{1}{2} \right) = q \left(\phi + \tfrac{1}{2} \right) \quad \text{and} \quad 
  \theta \left(\phi - \tfrac{1}{2} \right) = \theta \left(\phi+\tfrac{1}{2} \right),
\end{equation}
which follow from \eqref{eq:periodic_new}, and the deep water conditions \eqref{eq:deep} to derive the periodic Hilbert transform given by
\begin{equation}\label{eq:BI-qtheta}
\log{(q)}=\mathscr{H}[\theta](\phi)=\dashint_{-\frac{1}{2}}^{\frac{1}{2}}\theta(\phi') \cot{[\pi(\phi'-\phi)]} \de \phi'.
\end{equation}
In the above, $\dashint$ is the Cauchy principal-value integral. The above provides the crucial relationship between the components $q$ and $\theta$, and further details on the derivation of the boundary-integral relations can be found in chapter~{6} of \cite{vanden2010gravity}. 

Finally, the energy expression \eqref{eq:Energy} is also considered in terms of $(q,\theta)$. Noting that $x_{\phi}=q^{-1}\cos(\theta)$ and $y_{\phi}=q^{-1}\sin{\theta}$, we substitute $y=(F^2/2)(1-q^2)+Bq \theta_{\phi}$ from Bernoulli's equation to find
\begin{equation}\label{eq:energy-qtheta}
\E = \frac{1}{\Ehigh}\int_{-\frac{1}{2}}^{\frac{1}{2}} \Big[ \mathcal{G}_0(\phi)+B\mathcal{G}_1(\phi) +B^2\mathcal{G}_2(\phi)  \Big] \de \phi,
\end{equation}
\end{subequations}
where we have defined components
\begin{equation}
\begin{split}
\mathcal{G}_0(\phi) =& \frac{F^4}{8q}(1-q^2)(3 \cos{\theta}-2q-q^2 \cos{\theta}), \\
 \mathcal{G}_1(\phi)=& \frac{(1-\cos\theta)}{q}+\frac{F^2 \theta_{\phi}}{2}(2\cos{\theta}-q-q^2\cos{\theta}), \\
 \mathcal{G}_2(\phi)=& \frac{q \theta_{\phi}^2\cos{\theta}}{2}.
\end{split}
\end{equation}

In summary, the water-wave problem, as formulated for $q$ and $\theta$, involves the solution of equations \eqref{eq:Bern-qtheta}--\eqref{eq:energy-qtheta}. Note that the above sets of equations all involve the evaluation of $q$ and $\theta$ on the streamline $\psi = 0$.

\subsection{Analytic continuation}\label{sec:analcont}

\begin{figure}
\includegraphics[scale=1]{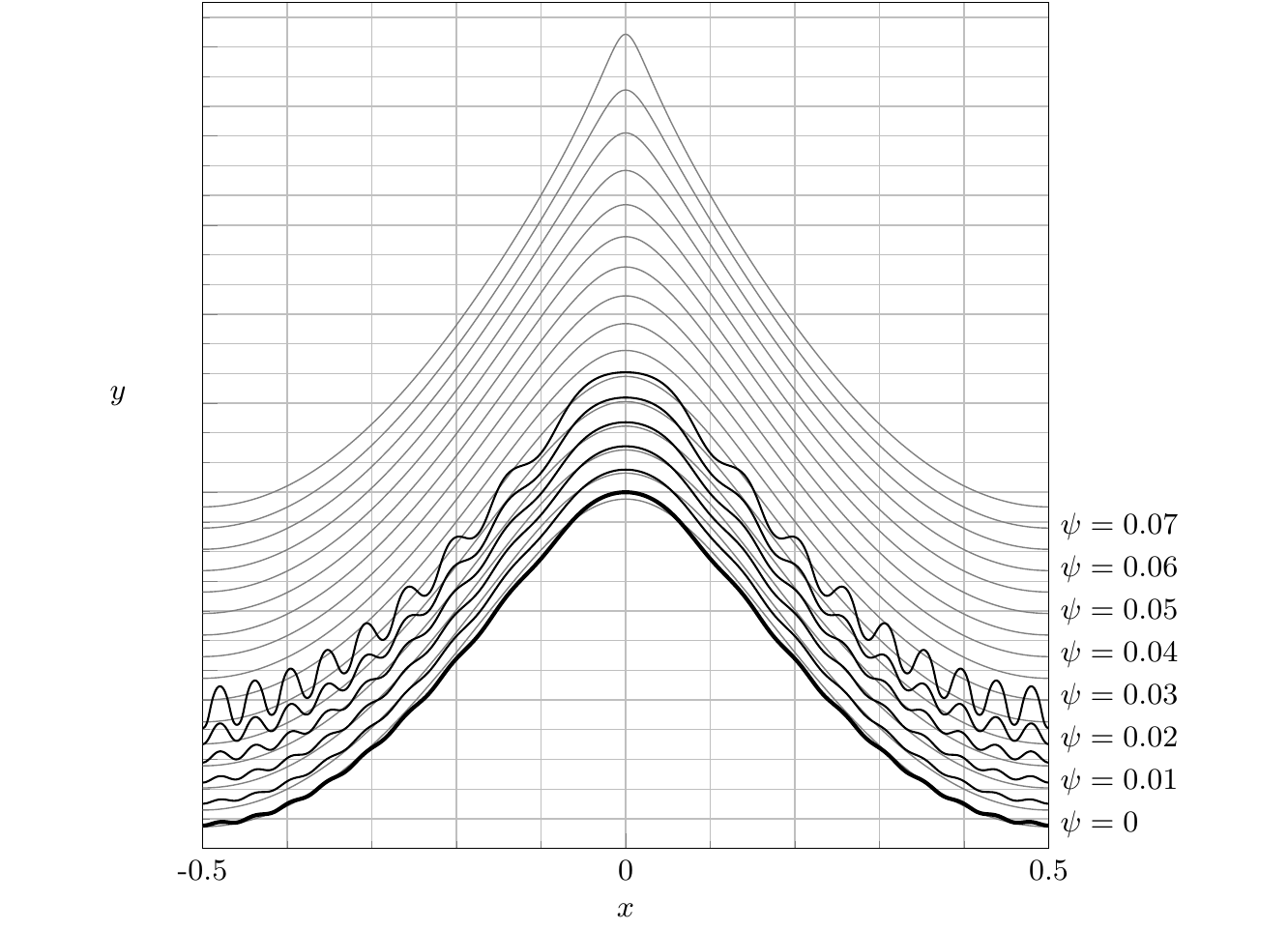}
\caption{\label{fig:numcont} The analytic continuation into the upper-half plane is calculated numerically for two solutions of equations \eqref{eq:MF1} with $\psi=0$. The first is a gravity-wave with $B=0$, $F=0.4104$, and $\E=0.3804$ (thin gray lines) and the second a gravity-capillary wave with $B=0.001$, $F=0.4188$, and $\E=0.3804$ (bold lines). The solutions with $\psi>0$ satisfy the analytically continued equations \eqref{eq:AC-system} and $\Re[x]$ vs $\Re[y]$ is shown. This image can be compared with figure~11 of \cite{longuet-higginsTheoryAlmosthighestWave1978}, which provides a streamline plot of the pure-gravity solution in the analytically continued plane.}
\end{figure}

\begin{figure}
\includegraphics[scale=1]{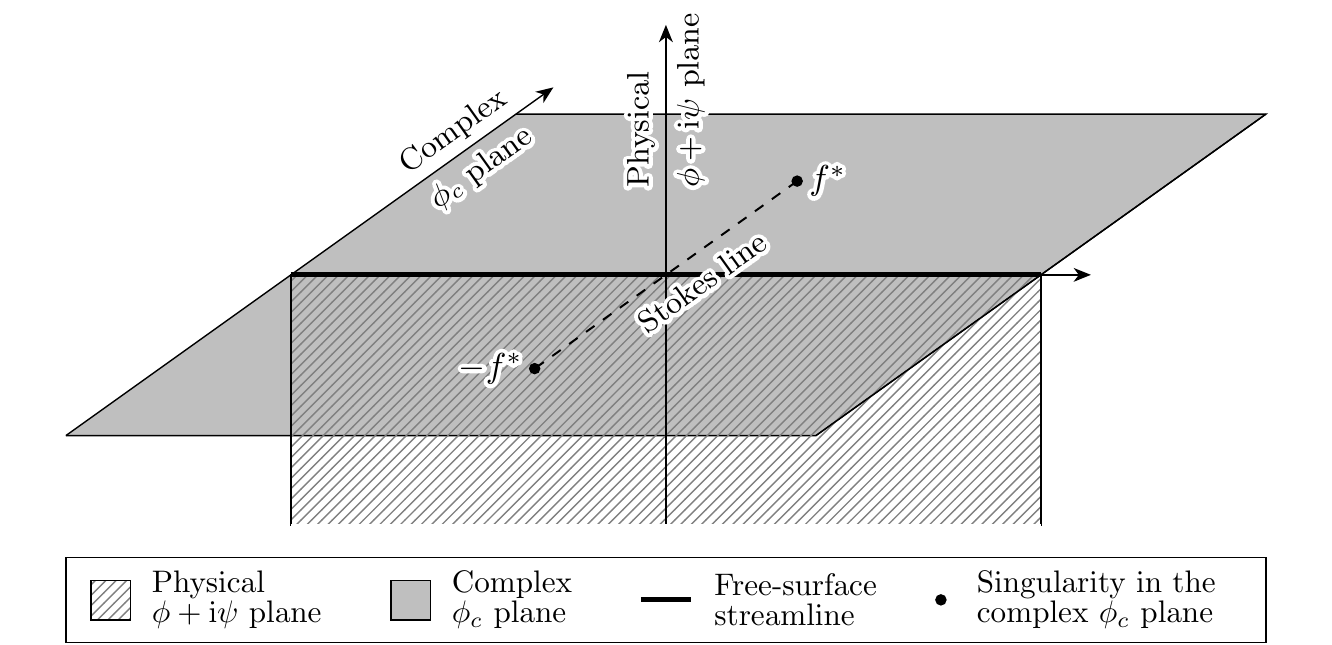}
\caption{\label{fig:phicont} A schematic of our analytic continuation procedure demonstrates the difference between the physical $\phi+ \i \psi$ plane and our complexified $\phi_c$ space. The location of the principle upper- and lower-half singularities at $f^*$ and $-f^*$ of the leading order flow field are shown by circles, and the main Stokes line from \S\ref{sec:expasymp} is shown dashed.}
\end{figure}

\noindent As we shall see, the exponential asymptotics procedure of \S\ref{sec:expasymp} will require the continuation of the free-surface solutions, $q(\phi + 0\im)$ and $\theta(\phi + 0\im)$, into the complex plane, where $\phi\in\mathbb{C}$. This free-surface continuation procedure is depicted in figure~\ref{fig:phicont}. Hence we shall analytically continue Bernoulli's equation \eqref{eq:Bern-qtheta} and the boundary-integral equation \eqref{eq:BI-qtheta} into the complex $\phi$-plane. 
The independent variable $\phi$ is complexified by considering $\phi \mapsto \phi_c\in\mathbb{C}$ and hence $q$ and $\theta$ are analytically continued. For convenience, we re-label $\phi_c$ as $f$. Thus Bernoulli's equation remains in an identical form to \eqref{eq:Bern-qtheta}, but with the variable $\phi$ replaced by $f$.

For the boundary-integral equation \eqref{eq:BI-qtheta}, we must consider the complexification of the Hilbert transform. Let us write
\begin{equation}
\label{eq:AC-hilb}
\mathscr{H}[\theta]=\widehat{\mathscr{H}}[\theta]-a\i \theta,
\end{equation}
where $\widehat{\mathscr{H}}[\theta]$ is the complex-valued Hilbert transform, 
\begin{equation*}
\label{eq:AC-cvht}
\widehat{\mathscr{H}}[\theta](f)=\int_{-\frac{1}{2}}^{\frac{1}{2}}\theta(\phi') \cot{[\pi(\phi'-f)]} \, \de{\phi'}.
\end{equation*}
Note that the integral above is only evaluated along the physical free-surface, parameterised in terms of $\phi^{\prime}$, where $\theta$ takes real-values. 

In \eqref{eq:AC-hilb}, we have also introduced the parameter, $a$, which is defined by 
 \begin{equation} \label{eq:mya}
    a=\left\{
                \begin{array}{ll}
                  +1 \ \ \text{for }\Im (f) >0,\\
 \\
                  -1 \ \  \text{for }\Im (f) <0. \\
                \end{array}
              \right.
  \end{equation}
When the Hilbert transform relationship is extended into the upper half-$f$-plane, $a = 1$, whereas $a = -1$ for continuation into the lower half-$f$-plane. The validity of \eqref{eq:AC-hilb} as a legitimate complexification of the Hilbert transform is verified by taking $\Im(f) \to 0$ on the right hand-side. Then $\widehat{\mathscr{H}}[\theta]$ yields a principal value integral and residue. The residue contribution changes sign between $\Im(f) \to 0^+$ and $\Im(f) \to 0^-$, yielding the constant $a$.

In summary, the governing equations for the analytically continued $q$ and $\theta$ values are given by
\begin{subequations} \label{eq:AC-system}
\begin{gather}
F^2 q^2 q'+ \sin{(\theta)}-Bq \big(q \theta' \big)'=0, \label{eq:AC-bern}\\
\log{(q)}+a{\mathrm{i}}\theta=\widehat{\mathscr{H}}[\theta], \label{eq:AC-bi}\\
\E = \frac{1}{\Ehigh}\int_{-\frac{1}{2}}^{\frac{1}{2}} \Big[ \mathcal{G}_0(\phi)+B\mathcal{G}_1(\phi) +B^2\mathcal{G}_2(\phi)  \Big] \de \phi, \label{eq:AC-en} \\
q(-\tfrac{1}{2})=q(\tfrac{1}{2})  ~~~ \text{and} ~~~ q^{\prime}(-\tfrac{1}{2})=q^{\prime}(\tfrac{1}{2}).\label{eq:periodicq}
\end{gather}
\end{subequations}
Note that while \eqref{eq:AC-bern} and \eqref{eq:AC-bi} are evaluated through complex $f$-space, the energy condition is most easily evaluated on the physical free-surface. Here and henceforth, we use primes ($'$) to denote differentiation in $f$. This system will be solved in \S\ref{sec:EA} with an expansion holding under the limit of $B \to 0$.

\section{A critical examination of the Longuet-Higgins (1963) theory}
\label{sec:LH}

\noindent In his 1963 work, \cite{longuet-higgins_1963_the_generation} proposed a theory for the generation of steady parasitic ripples by considering an asymptotic expansion for small surface tension such that a gravity wave was obtained at leading order. In \S{3} he wrote the following perturbative form for the solutions,
\begin{equation}
\label{eq:LHtrunc}
q(\phi,\psi)=q_0+ \bar{q}, \qquad \theta(\phi,\psi) = \theta_0 + \bar{\theta}, \qquad y(\phi,\psi)=y_0+\bar{y}, 
\end{equation}
with $y$ denoting the wave-height. All quantities are dimensional and functions of the potential, $\phi$, and stream function, $\psi$. Let us introduce the logarithm of the speed by $\tau=\log{(q/c)}$, where $c$ is the wave speed. In writing $\tau=\tau_0+\bar{\tau}$, this yields $q_0=c \e^{\tau_0}$ and $\bar{q}=q_0 \bar{\tau}$ for $\bar{\tau}$ assumed small.

The expression that Longuet-Higgins produced for the capillary ripples was [cf. equation ($5.18$) in \cite{longuet-higgins_1963_the_generation}]
\begin{subequations} \label{eq:LHsolution_sys}
\begin{equation}
\label{eq:LHsolution}
  \bar{\tau} - \i \bar{\theta} \sim F(\phi) \e^{ - \i c \alpha(\phi)/T'} \quad \text{for} \quad \phi>0,
\end{equation}
where the functional prefactor, $F(\phi)$, and exponent, $\alpha(\phi)$, are given by
\begin{gather}
\label{eq:LHprefactor}
  F(\phi) = 4 \i \exp{ \bigg( \i \int_{0}^{\phi} \frac{\partial \tau_0}{\partial \psi}\de \phi \bigg)} \int_{0}^{\infty}\bigg( \frac{\partial \tau_0}{\partial \psi} \cos{(\alpha c /T')}\bigg)\de{\phi}, \\
  \alpha(\phi)=\int_{0}^{\phi} \e^{\tau_0} \de{\phi}. \label{eq:LHprefactor_alpha}
\end{gather}
\end{subequations}
Here, $T'$ is the dimensional surface tension coefficient, assumed to be small. Note that $\alpha(\phi)$ involves integration of a real-valued $\e^{\tau_0}$ over real-valued $\phi$ and hence $\alpha$ is also real. 

One of the main contributions of our work is to provide an improvement on the above formulae, which contains a number of problems related to the capture of small ripples. The three most important issues are:
\begin{enumerate}[label=(\roman*),leftmargin=*, align = left, labelsep=\parindent, topsep=3pt, itemsep=2pt,itemindent=0pt ]
\item The functional form of the prefactor, $F(\phi)$, in \eqref{eq:LHprefactor_alpha} is incorrect; the form written above emerges as a consequence of certain asymptotic inconsistencies in the derivation. 
\item Longuet-Higgins correctly predicted that the capillary ripples would exhibit wavelengths scaling with $T'$, but in closer examination of \eqref{eq:LHsolution}, the expression predicts a wave-amplitude that is of $O(1)$ and independent of $T'$. We shall find that for small values of the surface tension, the wave-amplitude is exponentially small in $T'$ (indeed this should be clear from figure~\ref{fig:expscale}). 
\item The above formulation does not provide any restriction on the solution space (i.e. the existence of a solvability condition observed in the full numerical simulations). It particular, it does not capture any of the observed bifurcation structure seen in figure~\ref{fig:JSbifurc}. 
\end{enumerate}

\noindent Note that a portion of the \cite{longuet-higgins_1963_the_generation} work is devoted to studying the addition of viscosity and also incorporating the almost-highest wave theory of \cite{longuet-higginsTheoryAlmosthighestWave1977} into \eqref{eq:LHsolution_sys}. However, in the present authors' view, the treatment following \S{6} of the 1963 work becomes increasingly \emph{ad-hoc} and difficult to analyse in view of the fundamental issues with \eqref{eq:LHsolution_sys}.

We will now discuss the key issues (i) to (iii) above in detail.

\subsection{Asymptotic inconsistencies in Longuet-Higgins (1963)}

\noindent Numerical evidence was provided by \cite{Shelton2021onthe} (see figure~\ref{fig:expscale}) to demonstrate that, for those solutions exhibiting small-scale ripples on an underlying gravity wave, the amplitude of these parasitic ripples is exponentially-small as $T' \to 0$. Solutions that display such exponentially-small behaviour cannot be described purely by a typical Poincaré expansion which contains only algebraic powers of the small parameter; their description will instead appear {\it beyond-all-orders} of the standard Poincaré expansion.


We now review Longuet-Higgins' approach in our non-dimensional formulation (using the Bond number, $B$, and Froude number, $F$, in \eqref{eq:InFB} instead of $T'$ and $c$). We start with the integrated form of Bernoulli's equation from \eqref{eq:Inequation} given in terms of $y$ and the streamline-speed, $q$, as
\begin{equation}
\label{eq:LH2}
\frac{F^2}{2}q^2+y-B\frac{\partial q}{\partial \psi}=\text{constant},
\end{equation}
where the derivative in the $\psi$ direction can be converted to a derivative the $\phi$ direction via the Cauchy-Riemann equations. In his \S{3} Longuet-Higgins considered a perturbation $(\bar{y},\bar{q})$ about the gravity-wave $(y_0,q_0)$ with the truncations from \eqref{eq:LHtrunc} to find
\begin{equation}
\label{eq:LH3}
\frac{F^2}{2}(q_0^2+2q_0\bar{q}+\bar{q}^2)+(y_0+\bar{y})-B\bigg(\frac{\partial q_0}{\partial \psi} +\frac{\partial \bar{q}}{\partial \psi}\bigg)=\text{constant}.
\end{equation}
Here, the $O(1)$ terms, $\frac{F^2}{2} q_0^2+y_0=\text{const.}$, are satisfied exactly as this is the gravity-wave equation with solutions $(y_0,q_0)$. 
Thus we obtain
\begin{equation}
\label{eq:LH4}
 \underbrace{F^2q_0\bar{q}+\bar{y}-B \frac{\partial q_0}{\partial \psi}}_{O(B)} -\underbrace{B\frac{\partial \bar{q}}{\partial \psi}}_{O(B^2)}=-\underbrace{\frac{F^2\bar{q}^2}{2}}_{O(B^2)}.
\end{equation}
The asymptotic behaviour indicated by the under-braced quantities follows by making the standard assumption that the leading corrections, $\bar{y}$ and $\bar{q}$, are both of $O(B)$. Consequently, $\bar{q} \ll q_0$, and so Longuet-Higgins neglected the nonlinear term $\bar{q}^2$ on the right-hand side of this equation. However, the $O(B^2)$ term on the left-hand side was not neglected. This assumption, which appears in his equation~(5.1), is asymptotically inconsistent. In fact, this inconsistency is how Longuet-Higgins was able to produce approximations to an \emph{a priori} exponentially-small capillary ripple, since otherwise, all corrections are ripple-free and algebraic in $B$. 

The above asymptotic inconsistency is somewhat typical in early models of many exponential asymptotic problems. There are two (formally correct) methods to proceed with \eqref{eq:LH4}:

\begin{enumerate}[label=(\roman*),leftmargin=*, align = left, labelsep=\parindent, topsep=3pt, itemsep=2pt,itemindent=0pt ]
\item We may correctly treat $\bar{y}$ and $\bar{q}$ to both be of $O(B)$. The leading-order terms in equation \eqref{eq:LH4} are thus
\[
F^2q_0\bar{q}+\bar{y}-B \frac{\partial q_0}{\partial \psi}=0,
\]
and would yield the $O(B)$ capillary correction term. The procedure could be continued to quadratic orders of $B$ and higher, but the resultant perturbative solution would never yield an exponentially-small ripple. In essence, this is a derivation of the regular perturbative expansion and leads to the analysis of \S\ref{sec:EA}.
\item Alternatively, we may consider $\bar{y}$ and $\bar{q}$ to both scale as $\sim \e^{-\alpha /B}$, i.e. for solutions to be of WKB type. Since differentiation of this ansatz yields a factor of $1/B$, the dominant terms in equation \eqref{eq:LH4} change to
\[
\underbrace{F^2 q_0 \bar{q} - B \frac{\partial \bar{q}}{\partial \psi}}_{O(\e^{-\alpha/B})} = \underbrace{B q_0'}_{O(B)}.
\]
The form of the above equation would allow for the correct prediction of the WKB phase, $\alpha$, but not the correct prefactor (amplitude); this is on account of the fact the right-hand side is the result of a one-term truncation of the Poincar\'{e} expansion \eqref{eq:LHtrunc}. Instead, the correct procedure must involve additional terms of the regular expansion. In general, the right hand-side is of $O(B^N)$ with $N \to \infty$ as $B \to 0$. In order to derive the exponentially-small ripples we must optimally truncate with $N$ chosen carefully (\citealt{chapman_1998_exponential_asymptotics}).
\end{enumerate}
Longuet-Higgins had worked with the asymptotically inconsistent \eqref{eq:LH4}, with the right-hand side set to zero, and this was used to derive the solution \eqref{eq:LHsolution_sys}.

It will be shown in \S\ref{sec:SLS} that the ripples have the analytical behaviour
\begin{equation}
\label{eq:LH5}
q_{\text{exp}}(\phi)=\Lambda \mathcal{F}(\phi)\exp{\Big(-\frac{{\chi}(\phi)}{B}\Big)},
\end{equation}
where $\Lambda$ is a constant coefficient, $\mathcal{F}(\phi)$ is a functional prefactor, and $\chi(\phi)$ is the exponentially-small dependence of the solution, which is related to the quantity $\alpha(\phi)$. 
These components will be significantly different than those derived by Longuet-Higgins in \eqref{eq:LHsolution_sys}. In order to be correct, the above expression must be derived through optimal truncation of the standard asymptotic expansion, rather than using the one-term truncation in \eqref{eq:LHtrunc}. 

We note that it is still nevertheless possible to capture exponentially-small behaviour with the truncation \eqref{eq:LHtrunc} used by Longuet-Higgins. A comprehensive review of truncations of this type, for the case of free-surface flows, is given by \cite{trinh2017reduced} who, aided by the use of exponential asymptotics, discusses how the functional form of the exponentially-small waves changes when different truncations are made. The type utilised here by Longuet-Higgins in \eqref{eq:LHtrunc} is an $N=1$ truncation as only one term of the asymptotic series is included. While this truncation (if dealt with in an asymptotically consistent manner) can predict the correct exponentially-small scaling of the solution, the functional form of the prefactor and its magnitude [cf. \eqref{eq:LHprefactor}] will be incorrect.

\subsection{The choice of integration in the exponential argument}

\noindent We now discuss the second issue with Longuet-Higgins' analytical solution, which is that \eqref{eq:LHsolution_sys} predicts an $O(1)$ solution magnitude. For real values of $\phi$, $\alpha$ takes purely real values. Thus, as his solution contains $\e^{-\im c \alpha /T'}$, only a rapidly-oscillating waveform of wavelength $O(\epsilon)$ is predicted. The issue is not precisely one related to the functional form of the exponential argument, since modulo the scalings, it can be confirmed via our work that 
\[
-\frac{1}{B} \dd{\chi}{\phi} \ \propto \, -\frac{\im c}{T'} \dd{\alpha}{\phi}.
\] 

However, Longuet-Higgins restricts $\phi$ to take real values and forces the starting point of integration in $\alpha(\phi)$ to be at $\phi=0$. This is later matched to an \emph{ad-hoc} simplification near the crest of the wave. This misses a fundamental step in the determination of the parasitic ripples since, as we shall see, their existence is intimately connected with the singularities of ${\chi}^{\prime}{(\phi)}$ in the analytic continuation of the free-surface. In order to correctly resolve the Stokes phenomenon in \S\ref{sec:expasymp}, integration in our expression for $\chi$ must begin from such singularities, and results in a path of integration through the complex-valued domain. The final result produces a complex-valued $q_\text{exp}$, which is paired with a conjugate contribution to in order to produce a real-valued solution with both exponentially-small phase and amplitude.

\section{The expansion for small surface tension, $B$}
\label{sec:EA}

\noindent In the limit of $B \to 0$, we consider the traditional series expansions for $q$ and $\theta$, given by
\begin{equation}
\label{eq:expansions}
q=\sum_{n=0}^{\infty}B^nq_n \quad \text{and} \quad \theta=\sum_{n=0}^{\infty}B^n \theta_n.
\end{equation} 
These expansions will satisfy both Bernoulli's equation \eqref{eq:AC-bern} and the boundary-integral equation \eqref{eq:AC-bi} to each order in $B$. As noted in the discussion following \eqref{eq:energy-qtheta}, specifying $B$ and enforcing the energy constraint requires that $F$ be treated as an eigenvalue. Hence we also consider an expansion of the Froude number by 
\begin{equation}
\label{eq:expansionF}
F=\sum_{n=0}^{\infty}B^n F_n.
\end{equation} 

At leading order in \eqref{eq:AC-bern}, \eqref{eq:AC-bi}, and \eqref{eq:energy-qtheta} this results in the gravity-wave equations
\begin{subequations} \label{eq:O1}
\begin{gather}
F_0^2 q_0^2 \frac{\de q_0}{\de f}+ \sin{(\theta_0)}=0, \label{eq:O1-bern} \\
\log{(q_0)}+a{\mathrm{i}}\theta_0=\widehat{\mathscr{H}}[\theta_0], \label{eq:O1-bi} \\
\E=\frac{1}{\Ehigh}\int_{-\frac{1}{2}}^{\frac{1}{2}} \frac{F_0^4}{8q_0}(1-q_0^2)(3 \cos{\theta_0}-2q_0-q_0^2 \cos{\theta_0})\de \phi, \label{eq:O1-energy}
\end{gather}
\end{subequations}
where we remind the reader that $a = \pm 1$ via the choice of analytic continuation into the upper or lower half-planes, respectively [cf. \eqref{eq:mya}]. Here, the Hilbert transform in \eqref{eq:O1-bi} acts on the free-surface for which $f$ is real.
The energy, $\E$, is a specified $O(1)$ constant, which we take to be less than unity. 

At $O(B)$, we have for Bernoulli's equation,
\begin{subequations} \label{eq:OB}
\begin{equation}\label{eq:OB-bern}
F_0^2  q_0^2 \frac{\de q_1}{\de f}+ 2F_0^2q_0q_0^{\prime}q_1+2F_0F_1q_0^2q_0^{\prime} +\theta_1 \cos{\theta_0}-q_0\big(q_0 \theta_0^{\prime}\big)^{\prime}=0,
\end{equation}
for the boundary-integral equation, 
\begin{equation}\label{eq:OB-bi}
\frac{q_1}{q_0}+a{\mathrm{i}}\theta_1=\widehat{\mathscr{H}}[\theta_1],
\end{equation}
and finally for the energy constraint,
\begin{multline}\label{eq:OB-energy}
0 = \int_{-\frac{1}{2}}^{\frac{1}{2}} \bigg[\frac{(1-\cos{\theta_0})}{q_0} +\frac{F_0^2 \theta_0^{\prime}}{2}(2 \cos{\theta_0}-q_0-q_0^2\cos{\theta_0}) +\ldots \\
(3\cos{\theta_0}-2q_0-q_0^2\cos{\theta_0})\bigg(\frac{F_0^3F_1(1-q_0^2)}{2q_0}-\frac{F_0^4q_1}{8q_0}(1+q_0^2)\bigg)+\ldots\\
\frac{F_0^4(1-q_0^2)}{8q_0}(-3\theta_1\sin{\theta_0}-2q_1+q_0^2\theta_1\sin{\theta_0}-2q_0q_1\cos{\theta_0})\bigg]\de \phi.
\end{multline}
\end{subequations}

We now consider the $O(B^n)$ components of equations \eqref{eq:AC-bern} and \eqref{eq:AC-bi}.
The solutions of these, $q_n$, $\theta_n$, and $F_n$, are denoted the {\it late terms} of the asymptotic expansions \eqref{eq:expansions} and \eqref{eq:expansionF}.
An important feature of these solutions is that they diverge as $n \to \infty$. 
This is a consequence of the singularities in the leading order solutions, $q_0$ and $\theta_0$, which will be derived in \S \ref{sec:ACstokes}.
Evidently, the $O(B^n)$ equations will contain an unbounded number of terms as $n \to \infty$.
However, due the the divergent nature of the late-terms, only a few of these terms will influence the leading order solution as $n \to \infty$.

Starting with Bernoulli's equation \eqref{eq:AC-bern}, we retain the two leading orders in $n$, yielding
\begin{subequations}
\begin{multline} \label{eq:OBn-bern}
\biggl[ F_0^2\Big( q_0^2q_n^{\prime} +2q_0q_1q_{n-1}^{\prime}+2q_0q_0^{\prime}q_n+\ldots\Big)+2F_0F_1q_0^2q_{n-1}^{\prime}+2F_0F_nq_0^2q_0^{\prime}+\ldots \biggr] \\
+ \biggl[ \theta_n \cos{\theta_0}+\ldots \biggr] - \biggl[ q_0^2\theta_{n-1}^{\prime \prime}+2q_0q_1\theta_{n-2}^{\prime \prime}+q_0\theta_0^{\prime}q_{n-1}^{\prime}+q_0q_0^{\prime}\theta_{n-1}^{\prime}+\ldots\biggr] =0.
\end{multline}
At $O(B^n)$, we expand the logarithm in the boundary-integral equation \eqref{eq:AC-bi} in order to obtain
\begin{equation}
\label{eq:OBn-bi}
\frac{q_n}{q_0}-\frac{q_1q_{n-1}}{q_0^2}+\ldots +a\i \theta_n=\widehat{\mathscr{H}}[\theta_n].
\end{equation}
\end{subequations}

\section{On the singularities of the leading-order flow} \label{sec:ACstokes}

\noindent A crucial element of the exponential asymptotics analysis relies upon the understanding that the series \eqref{eq:expansions} will diverge on account of singularities (such as poles or branch points) in the analytic continuation of $q$ and $\theta$. More specifically, we shall find that the leading-order solution, $q_0$, which corresponds to the pure Stokes gravity wave via \eqref{eq:O1}, contains branch points in the complex plane. Since the determination of each subsequent order generally relies upon differentiating the previous, the result is that the order of the singularity increases as $n \to \infty$. This will be shown in \S\ref{sec:lateorders}.

On the assumption that the leading-order Stokes wave possesses a singularity in the complex plane, previously \cite{grant1973singularity} derived the local asymptotic behaviour using a dominant balance. That is, by considering the complex-velocity $\frac{\de f}{\de z}$ from equation \eqref{eq:cv}, he showed that near to a point $f^* \in \mathbb{C}$ directly `above' the wave-crest
\begin{equation}
\label{eq:grantsing}
\frac{\de f}{\de z} \sim (f-f^{*})^{\frac{1}{2}}.
\end{equation}

In the exponential asymptotics to follow, we require the singular behaviour of the individual components of $q_0$ and $\theta_0$. This is derived below, along with a discussion of the difference between Grant's singularity in $\de{f} / \de{z}$ and those of $(q_0, \theta_0)$. 

\subsection{Singularities in the analytic continuation of $q_0$ and $\theta_0$}
\label{sec:singqtheta}

\noindent The singular scaling of $q_0$ and $\theta_0$ is now considered. We let $f^*$ denote the `crest' singularity in the upper half-$f$-plane. We leave the constant, $a$, unspecified and take the limit of $f \rightarrow af^*$. First, it can be verified \emph{a posteriori} that as $f \to af^*$, $|\Im\theta_0| \to \infty$ and
\begin{equation*} 
\sin{\theta_0}=\frac{1}{2 \i}\Big[\e^{\i \theta_0}-\e^{-\i \theta_0}\Big] \sim \frac{a}{2\i }\e^{a\i \theta_0}.
\end{equation*}
We multiply Bernoulli's equation \eqref{eq:O1-bern} by $q_0$, and use the above scaling for $\sin{\theta_0}$ to find 
\begin{equation} \label{eq:bernmultq0}
F_0^2 q_0^3 \frac{\de q_0}{\de f}=-q_0\sin{\theta_0} \sim -\frac{a}{2\i }q_0 \e^{a\i \theta_0}.
\end{equation}
However, in taking the exponential of the boundary-integral equation \eqref{eq:O1-bi}, we have 
\begin{equation} \label{eq:Sing6}
q_0 \e^{a \i\theta_0} = \e^{\widehat{\mathscr{H}}[\theta_0]}.  
\end{equation}
Note that the complex Hilbert transform is applied to $\theta_0$ and integrated over the free-surface, where $\theta_0 = O(1)$. Thus $q_0 \e^{a \i\theta_0}$ is also of order unity and we conclude from \eqref{eq:bernmultq0} that $q^3_0q_0^{\prime}$ tends to a constant as $f \to af^*$. Integration then yields the following singular behaviour for $q_0$,
\begin{equation}
\label{eq:Sing8}
q_0 \sim c_a(f-af^*)^{\frac{1}{4}}.
\end{equation}
In addition, the scaling for $\e^{a\i \theta_0}$ is found from equation \eqref{eq:bernmultq0}, giving
\begin{equation}
\label{eq:Sing9}
\e^{a\i \theta_0} \sim \frac{-a\i F_0^2c_a^3}{2} (f-af^*)^{-\frac{1}{4}}.
\end{equation}
Combining these results for $q_0$ in \eqref{eq:Sing8} and $\theta_0$ in \eqref{eq:Sing9} gives the scaling for the complex velocity,
\begin{equation}
\label{eq:Sing10}
\frac{\de f}{\de z} \sim c_a\Big(\frac{-a \i F_0^2c_a^3}{2}\Big)^{-a}(f-af^*)^{\frac{a+1}{4}}.
\end{equation}
Note that $a=1$ recovers the same singular behaviour of \cite{grant1973singularity} in the upper half plane, shown in equation \eqref{eq:grantsing}.

\subsection{The apparent paradox of a singularity in the lower half-plane}

\noindent We see from equation \eqref{eq:Sing8} for $q_0$ that a singularity exists `within the fluid' in the lower half plane at $f=-f^{*}$.
This is in contrast to the regular behaviour near the same location provided by Grant's result.
Our apparent prediction of singular behaviour in the flow-field can readily be resolved by noting that this singularity is for the analytically continued variable, originally relabelled from $q_c \to q$ in \S\ref{sec:analcont}.
It is thus important to distinguish between the complexified and `physical' streamline speeds $q_c$ and $q_{\text{phys.}}$, and angles $\theta_c$ and $\theta_{\text{phys.}}$.
These physical variables are found by taking the magnitude and argument of the complex velocity $q_0 \e^{-\i \theta_0}$ as in equation \eqref{eq:Sing10}, which is regular for $a=-1$, yielding
\begin{equation}\label{eq:Sing11}
q_{\text{phys.}}=\Big\lvert q_c \e^{- \i \theta_c}\Big\rvert \quad \text{and} \quad \theta_{\text{phys.}}= \Arg{ \Big (q_c \e^{- \i \theta_c} \Big)}.
\end{equation}
Thus as $f \to -f^*$ these physical values are regular for the leading-order Stokes wave solution.
Only by recombining $q_0 \e^{-\i \theta_0}$ to find the physical values within the fluid have these singular terms cancelled out. 

\section{Exponential asymptotics} \label{sec:lateorders}

\noindent As we shall show in \S\ref{sec:expasymp}, the exponentially-small ripples are intimately connected with the later term divergence of the asymptotic series \eqref{eq:expansions}. In this section, we seek to characterise this divergence.

As we have noted in the previous section, the leading-order solution, $q_0$ and $\theta_0$, which represents a pure gravity wave, contains singularities at the points $f = af^{*}$, where $a=\pm 1$, (and further singularities on subsequent Riemann sheets---cf. \citealt{crew2016new}). Since later orders depend on successive differentiation of the previous orders, we intuit that as $n \to \infty$, the late terms of $q_n$ and $\theta_n$ diverge. In this limit of $n \to \infty$, the divergence can be described by a factorial-over-power ansatz of
\begin{equation}\label{eq:ansatz}
q_n \sim \frac{Q(f) \rmGamma(n+\gamma)}{\chi(f)^{n+\gamma}} \quad \text{and} \quad \theta_n \sim \frac{\Theta(f) \rmGamma(n+\gamma)}{\chi(f)^{n+\gamma}}.
\end{equation} 
Here, $Q$, $\Theta$, and $\chi$ are all functions of $f$, and $\gamma$ is assumed to be constant. Note that that more generally, there is a summation of contributions of factorial-over-power type---one for each singularity in $f \in\mathbb{C}$. Typically, the nearest singularities determine the leading-order divergence. Since the late terms are determined through a linear perturbative procedure, it is sufficient to consider the general ansatz \eqref{eq:ansatz} and add the appropriate contributions once the general forms of $Q$, $\Theta$, and $\chi$ are derived.

A consequence of enforcing the $O(B^n)$ energy condition with these solutions is that the Froude number, $F_n$, is determined as an eigenvalue of the system. Thus $F_n$ in \eqref{eq:expansionF} will also diverge in a similar factorial-over-power manner, given by
\begin{equation}\label{eq:ansatzf}
F_n \sim \frac{\delta(n) \rmGamma(n+\gamma)}{\Delta^{n+\gamma}}.
\end{equation} 
This unusual divergent form arises from satisfying the boundary conditions on the complete solution.
The presence of a divergent eigenvalue is a feature typically neglected in similar studies and it will not affect the solvability condition we shall derive in this work. However, we shall discuss some subtle considerations of this property in \S\ref{sec:Di}.

The $O(B^n)$ component of Bernoulli's equation \eqref{eq:OBn-bern} is a linear differential equation for $q_n$ and $\theta_n$, where terms containing the divergent Froude number, $F_n$, appear as a forcing term.
 We solve the homogeneous Bernoulli equation, for which the divergent eigenvalue $F_n$ does not appear. In the discussion of \S \ref{sec:Di} we provide a more detailed justification of why it is sufficient to neglect the divergent eigenvalue, $F_n$, and the $O(B^n)$ energy condition. This yields
\begin{subequations}
\begin{multline} \label{eq:OBn-noF}
F_0^2\Big( q_0^2q_n^{\prime} +2q_0q_1q_{n-1}^{\prime}+2q_0q_0^{\prime}q_n+\ldots \Big)+2F_0F_1q_0^2q_{n-1}^{\prime}+\ldots\\
+\theta_n \cos{\theta_0}-q_0^2\theta_{n-1}^{\prime \prime}-2q_0q_1\theta_{n-2}^{\prime \prime}-q_0\theta_0^{\prime}q_{n-1}^{\prime}-q_0q_0^{\prime}\theta_{n-1}^{\prime}+\ldots=0.
\end{multline}
In the above equation, we have explicitly written those terms that are necessary to correctly determine the leading and first order analysis of the late terms as $n \to \infty$. In particular, notice that if the ansatz \eqref{eq:ansatz} is differentiated once, then since $(n+\gamma)\rmGamma(n+\gamma) = \rmGamma(n+\gamma+1)$, the order in $n$ increases by one. Thus for example $q_{n-1}^{\prime} = O(q_n)$ as $n \to \infty$.

Next, we use the boundary-integral equation, \eqref{eq:OBn-bi} to substitute for $\theta_n$ in \eqref{eq:OBn-noF}. A key idea here, used in previous works on exponential asymptotics and water waves is that the term that involves the complex Hilbert transform, $\widehat{\mathscr{H}}[\theta_n]$, is evaluated on the real axis, and hence away from the singularities $f = af^*$. As a consequence, the contribution is exponentially subdominant to the left hand-side of \eqref{eq:OBn-bi} as $n \to \infty$. This idea of neglecting $\widehat{\mathscr{H}}[\theta_n]$ is a classic step in exponential asymptotics applications of many boundary-integral problems in interfacial flows (cf. \S{3} of \citealt{chapman_1999_on_the}, \S{5.3} of \citealt{trinh_2011_do_waveless} and \citealt{trinh2017reduced}) and can be rigorously justified in such cases (\citealt{tanveer2003analyticity}). 

With this in mind, we re-arrange the boundary-integral equation \eqref{eq:OBn-bi} to find
\begin{equation}
\label{eq:qnbi}
\theta_n \sim \frac{a \i q_n}{q_0} -\frac{a \i q_1 q_{n-1}}{q_0^2}+\ldots ~.
\end{equation} 
\end{subequations}
From this form, $\theta_{n-1}^{\prime \prime}$, $\theta_{n-2}^{\prime \prime}$, and $\theta_{n-1}^{\prime}$ are found in terms of $q_n$ and its derivatives.
Next, we substitute these into Bernoulli's equation \eqref{eq:OBn-noF} and consider the divergent ansatz \eqref{eq:ansatz}.
The leading order in $n$, which comes from the terms $q_n^{\prime}$ and $q_{n-1}^{\prime \prime}$, is seen to be of order $\rmGamma(n+\gamma+1)/\chi^{n+\gamma+1}$.
Dividing out by this divergence yields terms that are of $O(1)$, $O(1/n)$, and so on as $n \to \infty$.

Combining \eqref{eq:OBn-noF} and \eqref{eq:qnbi}, we obtain at leading order 
\begin{equation}
\label{eq:chieqn}
\chi^{\prime} (q_0 F_0^2 + a \i \chi^{\prime})=0.
\end{equation} 
We seek the non-trivial function $\chi$ that forces the divergence of the asymptotic expansion and hence takes the value of $\chi = 0$ at the singularities in $f$. Assuming that $\chi^{\prime} \neq 0$, we integrate to find
\begin{equation}
\label{eq:DE9}
\chi(f) =\chi_a(f) = a{\mathrm{i}}F_0^2\int_{af^*}^{f} q_0(f') \, \de{f'}.
\end{equation}
Here, we have chosen the starting point of integration to be the upper/lower-half singularity at $f = af^*$ where $a = \pm 1$. The function $\chi$, denoted the {\it singulant}, plays a pivotal role in the form of the exponentially-small terms and the associated Stokes smoothing procedure of \S\ref{sec:expasymp}. It will be convenient to distinguish the two singulants using the sub-index $a$. 

At the next order in Bernoulli's equation, $O(1/n)$, we use $\chi^{\prime}=a\i F_0^2 q_0$ and $\chi^{\prime \prime}=a\i F_0^2 q_0^{\prime}$ to find
\begin{equation}
\label{eq:Qeqn}
\frac{Q^{\prime}}{Q}=2\frac{q_0^{\prime}}{q_0}-a\i F_0^2q_1-2 a \i F_0F_1q_0+a\i \theta_{0}^{\prime}+\frac{a{\mathrm{i}}\cos{\theta_0}}{F_0^2 q_0^3}.
\end{equation}
Thus by integration, we find
\begin{equation}
\label{eq:Qsolution}
Q(f)= Q_a(f) = \Lambda_{a} q_0^2 \exp \bigg({a{\mathrm{i}}\theta_0+a \i \int_{0}^{f} \Big[ \frac{\cos{\theta_0}}{F_0^2 q_0^{3}} -F_0^2q_1-2F_0F_1q_0}\Big]\de{f'} \bigg).
\end{equation}
The starting point of integration has been chosen to be on the free-surface at $f=0$ for convenience. Other points may be chosen, which alters the value of the constant $\Lambda_a$. We note that this constant may take different values for $a=1$ and $a=-1$. Similarly, the form of $\Theta$ is found using \eqref{eq:qnbi} and thus 
\begin{equation}
\label{eq:qthetarelation}
  \Theta(f) = \Theta_a(f) = \frac{a\im Q_a(f)}{q_0(f)}.
\end{equation}

Substitution of this solution for $Q(f)$ into ansatz \eqref{eq:ansatz} then yields 
\begin{equation}
\label{eq:qsolution}
q_n(f) \sim \Lambda_{a} q_0^2 \exp \bigg({a{\mathrm{i}}\theta_0+a \i \int_{0}^{f} \Big[ \frac{\cos{\theta_0}}{F_0^2 q_0^{3}} -F_0^2q_1-2F_0F_1q_0}\Big] \, \de{f'} \bigg) \frac{\rmGamma(n+\gamma)}{\chi^{n+\gamma}},
\end{equation}
with $\chi$ given by \eqref{eq:DE9}. A similar form for $\theta_n$ may also be found by using the expression for $\Theta$ given in \eqref{eq:qthetarelation}. 

\subsection{Determination of $\gamma$ and $\Lambda$}\label{sec:innerlim}

\noindent At this point, we have determined the key components, $Q$, $\Theta$, and $\chi$, that appear in the factorial-over-power ansatz \eqref{eq:ansatz}. This leaves the value of the constants $\gamma$ and $\Lambda_a$. Note that our asymptotic series \eqref{eq:expansions} reorders as $f \to af^*$ (for which $q_0 = O(Bq_1)$ for instance) and the matched asymptotics procedure that results in investigating this limit yields $\gamma$ and $\Lambda_a$. 

In order to determine the constant $\gamma$, we take the limit $f \to af^*$ and match the order of the singularity of the divergent ansatz, valid for $n$ large, to the low-order behaviour. Setting $n=0$ in \eqref{eq:qsolution} and taking the limit of $f \to af^{*}$ yields
\begin{equation} \label{eq:gamma1}
q_n \Bigr\rvert_{n = 0} = O \left(\frac{q_0^2}{\chi^{\gamma}} \exp \bigg({a{\mathrm{i}}\theta_0+a \i\int_{0}^{f}\Big[ \frac{\cos{\theta_0}}{F_0^2 q_0^{3}} - F_0^2q_1-2F_0F_1q_0}\Big] \, \de{f'}\bigg)\right).
\end{equation}
From the scalings of $q_0$ and $\theta_0$ in \S\ref{sec:singqtheta}, and the scaling of $q_1$ in Appendix~\ref{app:q1} we find that
\begin{equation} \label{eq:gamma2}
\begin{gathered}
\chi^{\gamma} = O \left((f-af^*)^{\tfrac{5\gamma}{4}}\right), \\
q_0^2 \exp \bigg( a \i\int_{0}^{f} \Big[ \frac{\cos{\theta_0}}{F_0^2 q_0^{3}} -F_0^2q_1-2F_0F_1q_0 \Big]\de f \bigg) = O\left((f-af^*)^{\tfrac{5}{4}}\right).
\end{gathered}
\end{equation}
We substitute the above into \eqref{eq:gamma1} and match to $q_0 = O(f-af^*)^{1/4}$ to find 
\begin{equation}
\label{eq:gamma12}
\gamma=\frac{4}{5}.
\end{equation}

As is the case in many exponential asymptotic analyses, the determination of the constant prefactor, $\Lambda_a$, is often the most troublesome aspect of the procedure. For our purposes, it will be sufficient to know that $\Lambda_a$ is a non-zero constant, and can be determined via the solution of a numerical recursion relation. Specifically, it is found by matching the `inner' limit of $q_n$ from the divergent form \eqref{eq:qsolution} with the `outer' limit of the inner solution for $q$ near $f = af^*$. This analysis is performed in Appendix~\ref{sec:innermatching}, yielding
\begin{equation}
\label{eq:lambdaa}
\Lambda_a=-\frac{2 \i f^*}{F_0^2 c_a^4} \e^{-\mathcal{P}(af^*)}\bigg( \frac{4 a \i F_0^2 c_a}{5} \bigg)^{\tfrac{4}{5}} \lim_{n \to \infty} \frac{\hat{q}_n}{\rmGamma(n+\gamma)}.
\end{equation}
Here, $\hat{q}_n$ is the $n$th term of the outer-limit of an inner solution holding near $f=af^*$, and can be determined by recurrence relation \eqref{eq:IE10}. The constant $c_a$ is the prefactor of the singular scaling of $q_0$ from \eqref{eq:Sing9} while $\mathcal{P}(af^*)$ is given in \eqref{eq:Qinnerlim2}. We will not need to work with the precise value of $\Lambda_a$; however, later in \S{\ref{sec:ontheaxis}} and \S\ref{sec:SLS}, the fact that $\Lambda_{1}$ and $\Lambda_{-1}$ are complex conjugates will be crucial to obtain a real-valued solution on the free-surface. Since the prefactor, $\Lambda_a$, only has a scaling effect on the solutions (and is independent of $B$), it will be convenient to choose a specific value for visualisation purposes in \S{\ref{sec:Comparisons}}.

\subsection{The divergence along the free-surface} \label{sec:ontheaxis}

\noindent In order to capture the divergence of $q_n$ along the free-surface, $\text{Im}[f]=0$, we must include the effects of the two symmetrically-placed crest singularities indexed by $a = \pm 1$. We shall thus write
\[
\mathfrak{q}_{n}=q_n \rvert_{a=1}+q_n\rvert_{a=-1}.
\]
By the results of \S\ref{sec:innerconst}, the constants $\Lambda_{1}$ and $\Lambda_{-1}$ are the complex conjugates of one another. In regards to the two singulants, $\chi_1$ and $\chi_{-1}$, we may split the path of integration via
\begin{equation}
\label{eq:chionphi}
\chi_a(\phi)= a \i F_0^2 \bigg[ \int_{a f^{*}}^{0} +  \int_{0}^{\phi}\bigg] q_0(f') \, \de{f'}, 
\end{equation}
for $f = \phi$ along the real axis. As $q_0$ takes real values on the free-surface, $\text{Im}[f]=0$, the second integral above is seen to take purely imaginary values. By the Schwarz reflection principle, $q_0$ evaluated on the imaginary axis between $-af^*$ and $af^*$ is purely real and symmetric about the origin. Therefore the first integral on the right-hand side of \eqref{eq:chionphi} is purely real and takes the same value regardless of the choice of $a$. Thus, $\chi_{-1}$ and $\chi_{1}$ are also the complex-conjugate of one another on the free-surface.

Due to this behaviour of $\Lambda_a$ and $\chi_a$, we write
\begin{equation}
\label{eq:constandchi}
\Lambda_a=\lvert \Lambda_1 \rvert \e^{a \i \arg{\Lambda_1}} \quad \text{and} \quad \chi_a(\phi)= \lvert \chi_1(\phi) \rvert \e^{a \i \arg{\chi_1(\phi)}},
\end{equation}
which upon substitution into $\mathfrak{q}_{n}=q_n \rvert_{a=1}+q_n\rvert_{a=-1}$ yields
\begin{equation}
\label{eq:fullnaive}
\mathfrak{q}_n(\phi)= \frac{2 \lvert \Lambda_1 \rvert q_0^2 \rmGamma{(n+\gamma)}}{\lvert \chi_1(\phi)\rvert^{n+\gamma}} \cos{\bigg[\arg{\Lambda_1}-(n+\gamma)\arg{\chi_1}(\phi)+\theta_0+I(\phi) \bigg]},
\end{equation}
where we have defined
\begin{equation}
\label{eq:fullnaive2}
I(\phi)=\int_{0}^{\phi}  \Big( \frac{\cos{\theta_0}}{F_0^2 q_0^{3}} -F_0^2q_1-2F_0F_1q_0 \Big) \, \de{\phi'}.
\end{equation}
Thus the above form \eqref{eq:fullnaive} captures the real-valued divergence on the free-surface.

We have successfully derived an expression for the late term divergence on the axis in \eqref{eq:fullnaive} and off the axis in \eqref{eq:qsolution}.

\section{Stokes line smoothing}\label{sec:expasymp}

\noindent One of the key ideas of exponential asymptotics is that there exists a link between the factorial-over-power form of the divergences, given \eqref{eq:ansatz} and \eqref{eq:fullnaive}, and the exponentially-small terms we wish to derive. Following the work of \cite{dingle1973asymptotic}, {\it Stokes lines} are contours in the $f$-plane for which both
\begin{equation}
\label{eq:dinglecondition}
\Im[\chi_a(f)]=0 \quad \text{and} \quad \Re[\chi_a(f)] \geq 0.
\end{equation} 
Across and in the vicinity of these contours, exponentially small terms in the solution smoothly change in magnitude across a boundary layer. This is known as the {\it Stokes phenomenon}. In this section, we discuss the configuration of Stokes lines, and then perform the optimal truncation and Stokes-line-smoothing procedures needed to derive the exponentially small capillary ripples.

\subsection{Analysis of the Stokes lines}\label{sec:analysisstokes}

\noindent To find the Stokes lines for our problem, we apply conditions \eqref{eq:dinglecondition} to our expression for the singulant, $\chi_a$, given in \eqref{eq:DE9} as
\begin{equation*}
\label{eq:chi1}
\chi_a(f) =a{\mathrm{i}}F_0^2\int_{af^*}^{f} q_0(f') \de{f'}.
\end{equation*} 
Here, integration begins at the principal singularity, $f'=af^{*}$, that lies in the analytic continuation of the free-surface. Note that unlike many traditional studies in exponential asymptotics, the determination of the singulant function requires the leading-order solution, $q_0$, for which there does not exist a closed-form analytical solution. We will use numerical values of $q_0$ to evaluate the singulant, $\chi_a$.

\begin{figure}
\includegraphics[scale=1]{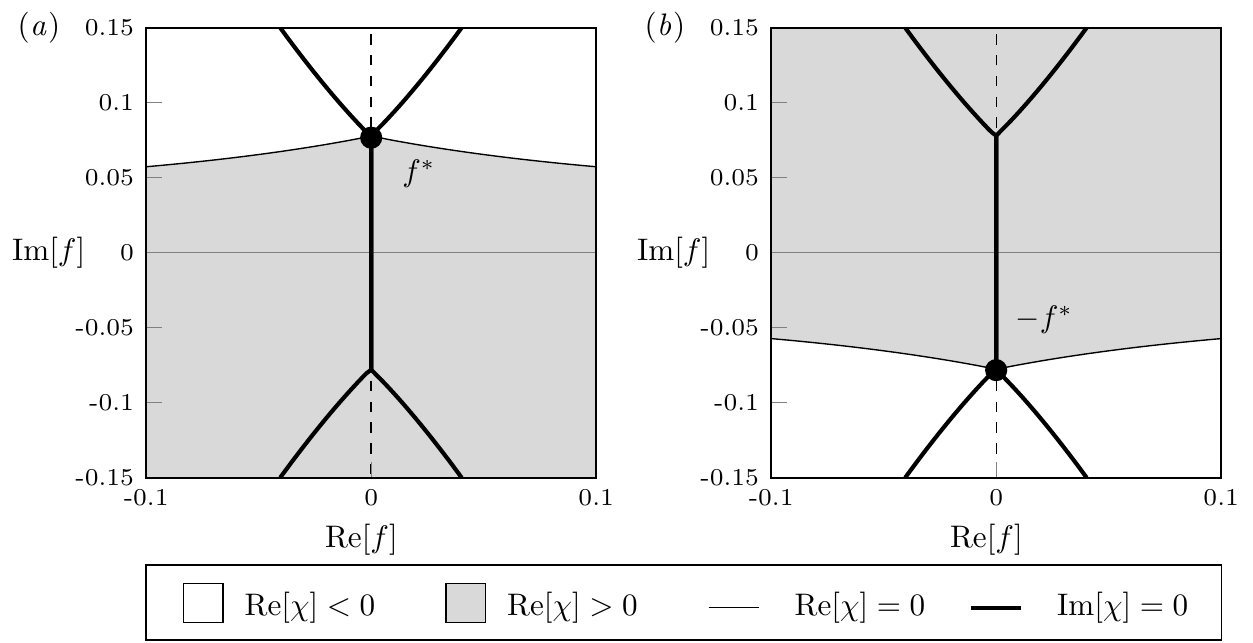}
\caption{\label{fig:singulant} Values of the singulant, $\chi_a$, are shown from the upper half singularity in $(a)$, with $a = 1$, and from the lower half singularity in $(b)$, with $a = -1$. The Stokes lines, which satisfy conditions \eqref{eq:dinglecondition}, are shown by the thick lines in the grey-shaded regions. This configuration corresponds to the energy $\E = 0.3804$ for which the upper half-plane singularity is at $f^* \approx 0.07776\i$. The chosen branch cuts for each of these singularities are shown by dashed lines.}
\end{figure}

The procedure is as follows. Given a fixed value of the energy, we obtain numerical values of $q_0$ and $\theta_0$ along the free-surface $\text{Im}[f]=0$ using the numerical computations of  \cite{Shelton2021onthe} or any standard procedure for calculating gravity Stokes waves (cf. \citealt{vanden1986steep}). Next, the analytic continuation method of \cite{crew2016new} is used to find $q_0$ and $\theta_0$ in the complex $f$-plane. Values for $\chi_a$ are then found across the domain by integrating $q_0$ along paths originating at either singularity. Graphs of the critical contours of $\text{Im}[\chi_a]$ and $\text{Re}[\chi_a]$ are given in figure~\ref{fig:singulant}
for the two choices of $a=1$ and $a=-1$. We see that there are two Stokes lines along the imaginary axis from $f=-f^*$ to $f=f^*$, one for $a=1$ and another for $a=-1$, which intersect with the free-surface at the wave-crest $\phi=0$.

Note that only the Stokes lines that intersect with the free-surface, $\text{Im}[f]=0$, are considered; other Stokes lines would indicate a switching-on or switching-off of exponentials in the general complex plane, but are not associated with the physical production of surface ripples.

\subsection{Optimal truncation}\label{sec:opttrunc}

\noindent In order to capture the exponentially-small components of the solution, which do not appear in the Poincaré series \eqref{eq:expansions}, we truncate the series at $n=N-1$ by considering
\begin{equation}\label{eq:OT1}
q= \underbrace{\sum_{n=0}^{N-1}B^n q_n}_{q_r} + \bar{q}, \quad 
\theta = \underbrace{\sum_{n=0}^{N-1}B^n \theta_n}_{\theta_r} + \bar{\theta}, \quad \text{and} \quad 
F = \underbrace{\sum_{n=0}^{N-1} B^n F_n}_{F_r} + \bar{F},
\end{equation} 
and thus we have introduced the notations of $q_r$, $\theta_r$, and $F_r$ for the truncated regular expansions of the solutions and eigenvalue. 

We will demonstrate that when truncated optimally at the point where two consecutive terms are of the same order, that is choose $N$ such that $\left\lvert B^N q_N \right\rvert \sim \left\lvert B^{N+1}q_{N+1}\right\rvert$, the remainders $\bar{q}$, $\bar{\theta}$, and $\bar{F}$ will be exponentially small.
This point of optimal truncation is given by
\begin{equation}\label{eq:OT1b}
N = \frac{\lvert\chi_a\rvert}{B} + \rho,
\end{equation} 
where $\rho\in[0, 1)$ is a bounded number to ensure $N$ is an integer. 

Substituting these into the boundary-integral equation \eqref{eq:AC-bi} yields a relationship between $\bar{\theta}$ and $\bar{q}$, given by
\begin{equation}\label{eq:OT2}
 \bar{\theta}=\frac{a \mathrm{i}\bar{q}}{q_r}-a \i \xi_{\text{int}}-a \i \widehat{\mathscr{H}}[\bar{\theta}]+O(\bar{q}^2).
\end{equation} 
Similarly we can insert the truncations \eqref{eq:OT1} into Bernoulli's equation \eqref{eq:AC-bern}.
This gives a second-order differential equation for $\bar{q}$ and $\bar{\theta}$. 
Upon substituting for $\bar{\theta}$ from \eqref{eq:OT2}, this is reduced down to an equation for $\bar{q}$ only.
Furthermore, we neglect the Hilbert transform of the remainder, $\widehat{\mathscr{H}}[\bar{\theta}]$, as this is anticipated to be exponentially subdominant.
This yields
\begin{multline}\label{eq:OT3}
\biggl[ a \i Bq_r \biggr] \bar{q}^{\prime \prime} 
+ \biggl[ -F_r^2q_r^2-a \i Bq_r^{\prime}+Bq_r\theta_r^{\prime}-a \i Bq_r \xi_{{\text{int}}}^{\prime}\biggr] \bar{q}^{\prime} + \bigg[ -\frac{a \i \cos{\theta_r}}{q_r}-2F_r^2q_rq_r^{\prime}+\frac{a \i B(q_r^{\prime})^2}{q_r} \\
+Bq_r^{\prime}\theta_r^{\prime} 
 - a \i Bq_r^{\prime \prime}+2Bq_r\theta_r^{\prime \prime}-
a \i B q_r^{\prime}\xi_{\text{int}}^{\prime}-2a \i Bq_r \xi_{\text{int}}^{\prime \prime} \biggr] \bar{q}
-2F_rq_r^2q_r^{\prime}\bar{F} = \mathcal{R} + O(\bar{q}^2).
\end{multline}
This is a second order differential equation for $\bar{q}$, in which the forcing terms on the right hand side are of $O(B^N)$. A similar equation was derived by \cite{trinh2017reduced} for the low-Froude limit of gravity waves. Here, we have introduced the forcing terms $\xi_{\text{int}}$ and $\xi_{\text{bern}}$ arising from the Poincaré expansion in the boundary-integral and Bernoulli's equations as
\begin{subequations} \label{eq:OT2a}
\begin{align}
\xi_{\text{int}}&=\widehat{\mathscr{H}}[\theta_r]-a \i \theta_r -\log{q_r}, \label{eq:OT2b}\\
\xi_{\text{bern}}&=F_r^2q_r^2q_r^{\prime}+\sin{(\theta_r)}-B(q_r^2\theta_r^{\prime \prime}+\theta_r^{\prime}q_r^{\prime}q_r), \label{eq:OT2c} \\
\mathcal{R} &= \xi_{\text{bern}}-a \i \cos{\theta_r}\xi_{\text{int}}+a \i Bq_rq_r^{\prime}\xi_{\text{int}}^{\prime}+a \i Bq_r^2\xi_{\text{int}}^{\prime \prime}. \label{eq:OT2_R} 
\end{align}
\end{subequations}
Due to the truncation at $n=N-1$, the equation \eqref{eq:OT3} is satisfied exactly for every order up to and including $B^{N-1}$ since $\xi_{\text{int}} = O(B^N)$ and $\xi_{\text{bern}} = O(B^N)$.

\subsection{Stokes line smoothing}
\label{sec:SLS}

\noindent We now seek a closed-form asymptotic expression for $\bar{q}$ and the terms switched-on across Stokes lines. We start with the homogeneous form of equation \eqref{eq:OT3}, in which the terms on the right-hand side and $\bar{F}$ are neglected. Following the exponential asymptotics methodology established in \emph{e.g.} \S{4} of \cite{chapman2006exponential}, we note that the homogeneous problem has solutions of the form, 
\begin{equation}
  \bar{q}_\text{homog.} \sim Q_a(f) \exp\Big(-\frac{\chi_a(f)}{B}\Big),
\end{equation}
where $\chi_a(f)$ and $Q_a(f)$ satisfy those same equations as found for the late-term ansatz via \eqref{eq:chieqn} and \eqref{eq:Qeqn}. To observe the Stokes phenomenon and the switching of exponentials, we now include the forcing terms on the right-hand side of equation \eqref{eq:OT3} for $\bar{q}$.
We consider a solution of the form
\begin{equation}
\label{eq:SS1}
\bar{q}(f)=A_a(f)Q_a(f)\exp\Big(-\frac{\chi_a(f)}{B}\Big),
\end{equation} 
where the Stokes multiplier $A_a(f)$ is introduced to capture the switching behaviour that occurs across the Stokes lines.
When the truncation point, $N$, is chosen optimally as in \eqref{eq:OT1b}, $\bar{q}$ will be seen to be exponentially small and will change in magnitude across the lines where $\Im[\chi_a]=0$ and $\Re[\chi_a]\geq 0$.

The algebra for this procedure follows very similarly to \emph{e.g.} \cite{chapman_1998_exponential_asymptotics,chapman2006exponential,trinh2017reduced}. Thus, when the exponential form of \eqref{eq:SS1} for $\bar{q}$ is substituted into \eqref{eq:OT3}, the dominant balance at leading-order is identically satisfied by our choice of $\chi$ determined in \eqref{eq:chieqn}. The first non-trivial balance occurs at $O(\e^{-\chi/B})$ which also involves the forcing terms on the right-hand side. We extract the $O(B^N)$ terms from $\mathcal{R}$ in \eqref{eq:OT2_R}, and this yields $\mathcal{R} \sim -q_0^2\theta_{N-1}^{\prime \prime}B^N$. The governing equation for $A_a$ is then given by 
\begin{equation}
\label{eq:SS5}
\left[F_0^2q_0^2  Q_a \e^{-\frac{\chi_a}{B}} \right] \dd{A_a}{f} \sim   -a \i q_0 q_{N-1}^{\prime \prime}B^N,
\end{equation} 
where we have used $\theta_{N-1}^{\prime \prime} \sim a \i q_0^{-1} q_{N-1}^{\prime \prime}$ from the boundary-integral equation \eqref{eq:qnbi}.

By substituting in the factorial-over-power form for $q_{N-1}^{\prime \prime}$ from \eqref{eq:ansatz}, and using the chain rule to change differentiation to be in terms of $\chi_a$, we find
\begin{equation}
\label{eq:SS6}
\frac{\de{A_a}}{\de{\chi_a}}=\frac{B^N \e^{\chi_a/B}\rmGamma(N+1+\gamma)}{\chi_a^{N+1+\gamma}}.
\end{equation} 
This is now of an equivalent form to that found by \cite{chapman2006exponential} for the low-Froude limit of gravity waves [cf. their equation (4.4)]. In brief, the procedure is as follows. First, we write $\chi_a = r_a \e^{\im \vartheta_a}$ and truncate optimally via \eqref{eq:OT1b} with $N = r_a/B + \rho$. Examination of the differential equation \eqref{eq:SS6} shows that there exists a boundary layer at $\vartheta_a = 0$ and indeed this is the anticipated Stokes line where $\Im[\chi_a]=0$. The appropriate inner variable near the Stokes line is $\vartheta_a=B^{1/2}\bar{\vartheta_a}$ and \eqref{eq:SS6} can then be integrated to show
\begin{equation}
\label{eq:Asolution}
A_a(f)=C_a + \frac{\sqrt{2 \pi} \i}{B^{\gamma}} \int_{-\infty}^{\bar{\vartheta}_a \sqrt{r_a}} \exp{(-t^2 / 2)} \, \de t,
\end{equation}
where $C_a$ is constant. Taking the outer limit of $\bar{\vartheta} \to \infty$, we then see that across the Stokes line, there is a jump of
\begin{equation}
\label{eq:SS7}
A_a(\vartheta_a \to 0-)-A_a(\vartheta_a \to 0+)=\frac{2\pi \i }{B^{\gamma}}.
\end{equation} 

\begin{figure}
\includegraphics[scale=1]{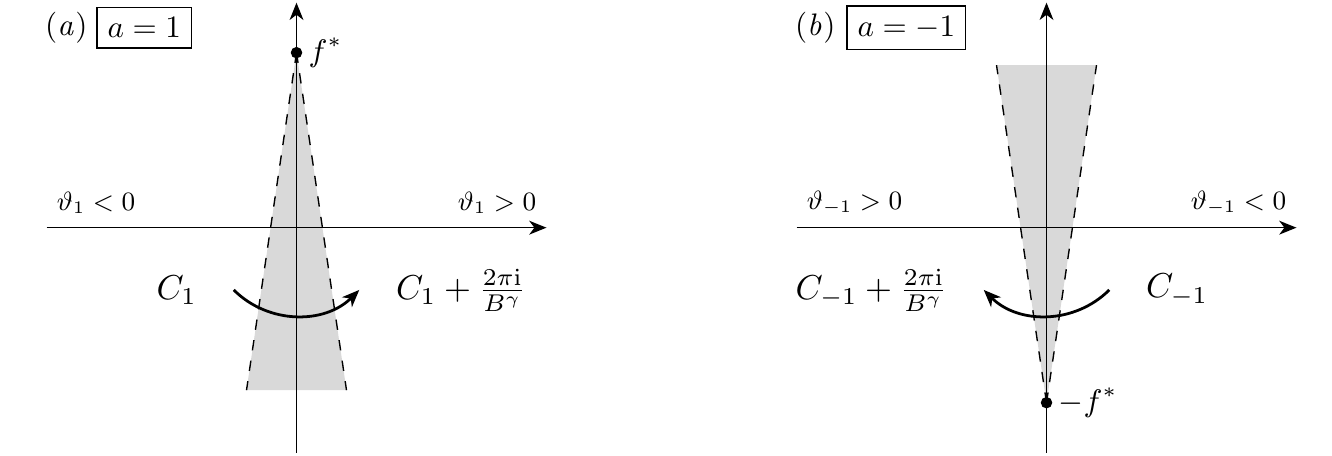}
\caption{\label{fig:smoothingUL} The Stokes smoothing procedure is visualised for $a=1$ in $(a)$ and for $a=-1$ in $(b)$.}
\end{figure}

As it concerns the relationship between Stokes-line contributions from $f = f^*$ and $f = -f^*$, note that as $\chi_{1}$ is the complex-conjugate of $\chi_{-1}$, we have $\vartheta_{1}=-\vartheta_{-1}$. Thus we anticipate that $C_1$ switches to $C_1 + 2\pi\im/B^\gamma$ as one proceeds from left-to-right across the Stokes line from $f = f^*$. This is shown in figure~\ref{fig:smoothingUL}(a). On the other hand, $C_{-1}$ switches to $C_{-1} + 2\pi\im/B^\gamma$ proceeding from right-to-left across the Stokes line from $f = -f^*$. This is shown in figure~\ref{fig:smoothingUL}(b). We emphasise that the above Stokes smoothing procedure only provides the local change of the prefactor, $A_a$, across the Stokes line. Determination of the constant, $C_a$, will follow from imposition of the boundary-conditions.

Returning now to \eqref{eq:SS1}, we write the leading-order exponentials on the axis, $\text{Im}[f]=0$, via $\bar{\mathfrak{q}}=\bar{q}\rvert_{a=1}+\bar{q}\rvert_{a=-1}$, either as an inner solution
\begin{subequations}
\begin{equation}
\label{eq:SStotal}
\bar{\mathfrak{q}}(\phi)=A_1(\phi)Q_1(\phi)\exp\bigg(-\frac{\chi_1(\phi)}{B}\bigg)+A_{-1}(\phi)Q_{-1}(\phi)\exp\bigg(-\frac{\chi_{-1}(\phi)}{B}\bigg),
\end{equation} 
for which $A(\phi)$ is given by \eqref{eq:Asolution}, or as an outer-solution by
\begin{equation}
\label{eq:SS8}
\bar{\mathfrak{q}}(\phi) \sim
\left\{\begin{aligned}
 C_1 \Big(Q_{a} \e^{-\frac{\chi_{a}}{B}}\Big)\biggr\rvert_{a=1} + \biggl\{ C_{-1}+\frac{2 \pi \i}{B^\gamma}\bigg\} \Big(Q_{a} \e^{-\frac{\chi_{a}}{B}}\Big)\biggr\rvert_{a=-1} \qquad \text{for $\phi<0$},\\
 \biggl\{ C_1 + \frac{2 \pi \i}{B^\gamma}\biggr\} \Big(Q_{a} \e^{-\frac{\chi_{a}}{B}}\Big) \biggr\rvert_{a=1}
 +C_{-1} \Big(Q_{a} \e^{-\frac{\chi_{a}}{B}}\Big) \biggr\rvert_{a=1} \qquad \text{for $\phi>0$}.
\end{aligned}\right. 
\end{equation}
\end{subequations}

In \eqref{eq:SS8}, the constants, $C_1$ and $C_{-1}$, will be determined by enforcing periodicity on $\bar{\mathfrak{q}}$ and $\bar{\mathfrak{q}}^{\prime}$, as given by
\begin{equation}
\label{eq:SS9}
\bar{\mathfrak{q}}(-1/2) =\bar{\mathfrak{q}}(1/2)  \quad \text{and} \quad \bar{\mathfrak{q}}^{\prime}(-1/2) =\bar{\mathfrak{q}}^{\prime}(1/2).
\end{equation} 
The second relation above arose by evaluating the derivative of periodicity condition \eqref{eq:periodicqth} at $\phi=0$.
In writing $C_1=C_{1}^R + \i C_{1}^I$ and $C_{-1}=C_{-1}^R+ \i C_{-1}^I$, we have four unknowns balancing the four equations from the real and imaginary parts of \eqref{eq:SS9}. 
Using $\Lambda_a=\lvert \Lambda_1 \rvert \e^{a \i \arg{\Lambda_1}}$ from equation \eqref{eq:constandchi} and $\chi_a=\Re[\chi_1]+ a \i \Im[\chi_{1}]$ then yields the solutions
\begin{equation}\label{eq:SS10}
\begin{split}
C_{1}^I=&-\frac{\pi}{B^{\gamma}}, \qquad C_{1}^R=-\frac{\pi}{B^{\gamma}}\frac{\cos{[G(1/2)]}}{\sin{[G(1/2)]}}, \\
C_{-1}^I=&-\frac{\pi}{B^{\gamma}}, \qquad C_{-1}^R=-\frac{\pi}{B^{\gamma}}\frac{\cos{[G(1/2])}}{\sin{[G(1/2)]}},
\end{split}
\end{equation}
where
\begin{equation}
\label{eq:SS11}
G(\phi)=\theta_0(\phi)+\int_{0}^{\phi}\bigg[ \frac{\cos{\theta_0}}{F_0^2 q_0^3}-F_0^2q_1-2F_0F_1q_0-\frac{F_0^2 q_0}{B}\bigg] \, \de{\phi}.
\end{equation} 
Solutions are not possible when $\sin[G(1/2)]=0$, from which we obtain the following discrete set of values of $B$,
\begin{equation}
\label{eq:SS11b}
B_n = \frac{F_0^2 \int_{0}^{1/2} q_0 \de \phi}{\theta_0(1/2)+\int_0^{1/2}\big[ \frac{\cos{\theta_0}}{F_0^2 q_0^3}-F_0^2q_1-2F_0F_1q_0\big] \de \phi + n \pi} \quad \text{for} \quad n \in \mathbb{Z}^+.
\end{equation} 
The above formula \eqref{eq:SS11b} provides the crucial eigenvalue condition for the non-existence of solutions. Recall that the ``parameters" in this formula, e.g. $\{F_0, F_1, q_0, q_1, \theta_0\}$, are dependent on the chosen energy, $\E$, in \eqref{eq:AC-en}. Note that $\theta_0(1/2) = 0$ and in addition, only solutions with positive integer values of $n$ correspond to positive values of the Bond number. Thus for instance, it is predicted that solutions do not exist at a countably infinite set of discrete values,
\begin{equation}
  B_1(\E) > B_2(\E) > B_3(\E) > \ldots > B_n(\E) > \ldots > 0.
\end{equation}
In the next section, we will show that these values of $B$ are associated with points between adjacent `fingers' of solutions in the bifurcation diagram. 

Substitution of \eqref{eq:SS10} for $C_1$ and $C_{-1}$ into \eqref{eq:SS8} then gives a real-valued solution on the free-surface. Firstly for $\phi<0$, we have
\begin{subequations} \label{eq:SS12}
\begin{equation}
\bar{\mathfrak{q}}(\phi) =
 -\frac{2\pi}{B^\gamma}\lvert \Lambda_1 \rvert  q_0^2 \e^{-\tfrac{\Re[\chi_1]}{B}} \bigg[ \frac{\cos{(G(1/2))}}{\sin (G(1/2))} \cos \big[\arg{\Lambda_1}+G(\phi)\big]- \sin \big[\arg{\Lambda_1}+G(\phi)\big] \bigg],
 \end{equation}
 while for values on the the positive real axis $\phi>0$, 
 \begin{equation}
 \bar{\mathfrak{q}}(\phi) = -\frac{2\pi}{B^\gamma}\lvert \Lambda_1 \rvert  q_0^2 \e^{-\tfrac{\Re[\chi_1]}{B}} \bigg[ \frac{\cos{(G(1/2))}}{\sin (G(1/2))} \cos \big[\arg{\Lambda_1}+G(\phi)\big]+\sin \big[\arg{\Lambda_1}+G(\phi)\big] \bigg].
\end{equation}
\end{subequations}
Note that the above forms for $\bar{\mathfrak{q}}$ are valid away from the boundary layer surrounding the Stokes line at  $\phi = 0$.

\section{Numerical comparisons with the full water-wave model}\label{sec:Comparisons}

\noindent We will now compare the asymptotic results of \S\ref{sec:SLS} to the numerical solutions of the fully nonlinear equations \eqref{eq:laplace}--\eqref{eq:deep} found by \cite{Shelton2021onthe}. These numerical solutions were calculated using a spectral method on a domain, $\phi$, uniformly discretised with $N=1024$ points [cf. \S{4} of \citealt{Shelton2021onthe} for details]. 

\subsection{Finding values for our analytical solution}

\begin{figure}
\includegraphics[scale=1]{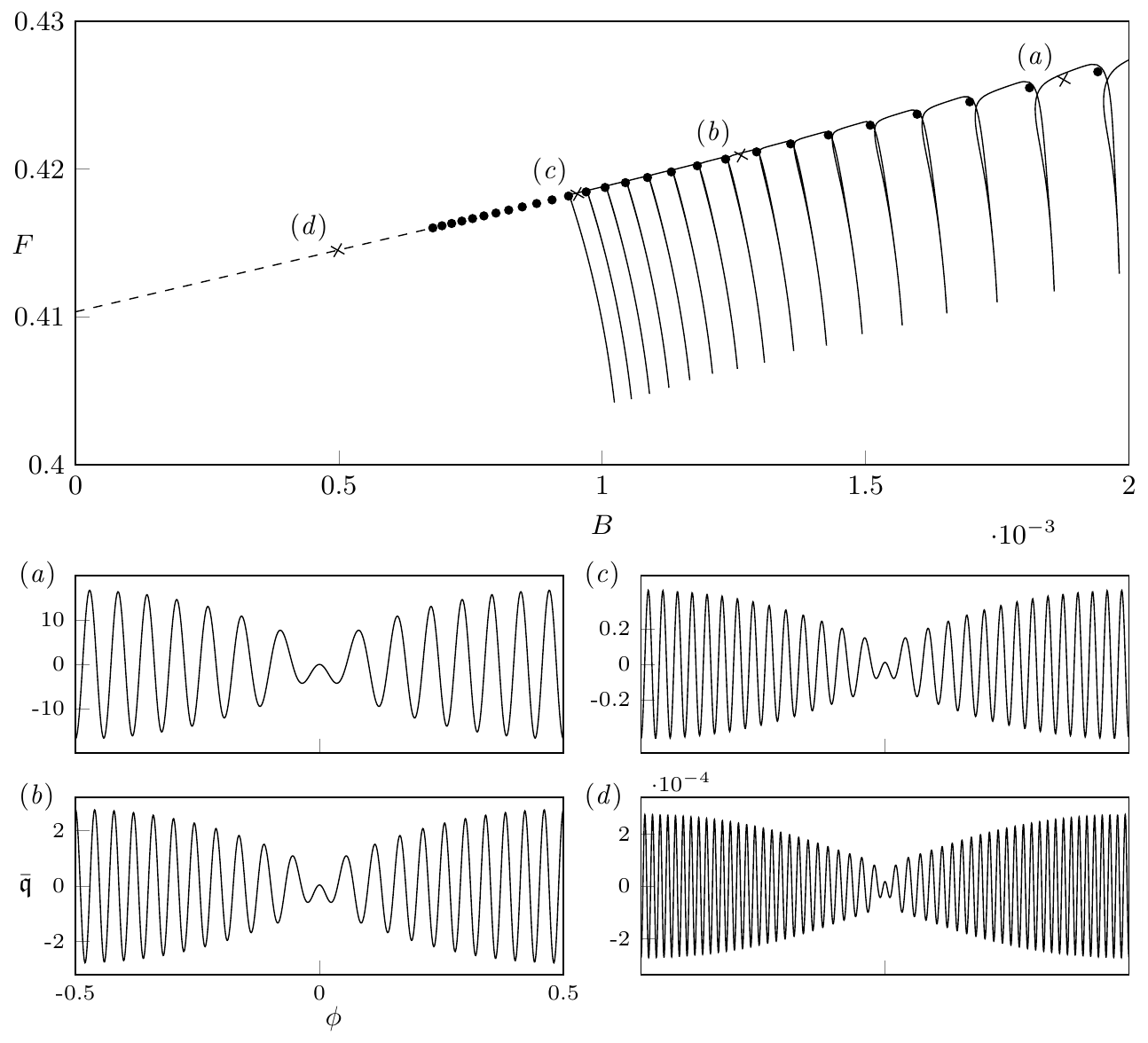}
\caption{\label{fig:solvability} A comparison between the numerical solution branches of  \cite{Shelton2021onthe} (shown solid) and the analytical approximations of $B_n$ from \eqref{eq:SS11b} (shown as black circles). The insets (a)--(d). These smaller insets show the exponentially-small ripples, $\bar{\mathfrak{q}}$, from equation \eqref{eq:SS12} for the four locations of $B=0.001876$, $B=0.001264$, $B=0.0009527$, and $B=0.0004978$ (shown as crosses in the main inset). The solutions are all computed at $\E = 0.3804$. A value of $\lvert \Lambda_a \rvert=1$ has been used for the constant prefactor.}
\end{figure}

\noindent To obtain precise values for our analytical solution, $\bar{\mathfrak{q}}$, across the domain, we use the form given in equation \eqref{eq:SStotal}. This form includes the local change across the boundary layer at $\phi=0$ and requires known values of $q_0$, $\theta_0$, $F_0$, $q_1$, and $F_1$ for a specified value of the energy, $\E$. 

In order to calculate values for these nonlinear solutions, we employ Newton iteration on the $O(1)$ and $O(B)$ equations \eqref{eq:O1} and \eqref{eq:OB} with an even discretisation of the domain, $\phi$. With these values known, the three components of $\bar{\mathfrak{q}}$, the Stokes-prefactor, $A_a(\phi)$, the functional pre-factor, $Q_a(\phi)$, and the singulant, $\chi_a(\phi)$, may then be calculated individually with a specified value of $B$:
\begin{enumerate}[label=(\roman*),leftmargin=*, align = left, labelsep=\parindent, topsep=3pt, itemsep=2pt,itemindent=0pt ]
\item For $Q_a(\phi)$ given in equation \eqref{eq:Qsolution}, we take the previously-computed values for $\theta_0$, $q_0$, $F_0$, $q_1$, and $F_1$ and employ numerical integration across the domain. As noted in \S{\ref{sec:innerlim}}, it is convenient to choose a value of $\lvert \Lambda_a \rvert$ in order to facilitate visualisation of the ripples. In figures~\ref{fig:solvability} and \ref{fig:onefinger}, we plot $\bar{\mathfrak{q}}$ with $\lvert \Lambda_a \rvert=1$. In figure \ref{fig:ongravity}, in order to compare between asymptotic and numerical solutions, we have chosen $\lvert \Lambda_a \rvert=0.006$, which is estimated by numerical fitting. It can be verified that fitting to other fingers changes the constant by only a small amount.
\item To determine $\chi_a(\phi)$, we split the range of integration as in \eqref{eq:chionphi}. This allows for $\text{Re}[\chi_a]$ to be calculated by integrating $q_0$ through the complex-valued domain from the singularity at $f=af^*$ to the wave crest at $f=0$. Next, $\text{Im}[\chi_a]$ is found by integrating $q_0$ over the free-surface from $f=0$ to $f=\phi$. Values for the integrand, $q_0$, are found with the analytic continuation method from \cite{crew2016new} described in \S{\ref{sec:analysisstokes}}.
\item To find the Stokes prefactor, $A_a(\phi)$, from equation \eqref{eq:Asolution}, the upper limit of the integral is determined by using $r_a=\lvert \chi_a \rvert$ and $\bar{\vartheta}=\frac{\arg{\chi_a}}{B}$ from the known values of $\chi_a$. The integral is then calculated with known values of the error function. The constants $C_a$ are then found by calculating $G(1/2)$ from \eqref{eq:SS11}.
\end{enumerate}
This process yields values for our exponentially-small component of the solution, $\bar{\mathfrak{q}}$, for specified values of $B$ and $\E$. The values of $B_n$ where the solvability condition fails from equation \eqref{eq:SS11b} are also found with the same method used for $Q_a$ above.

\subsection{Comparisons}

\noindent We begin by comparing the values of $B_n$ (where the solvability condition fails) to the ($B,F$) bifurcation space computed numerically by \cite{Shelton2021onthe}. In taking the same value of the energy, $\E=0.3804$, we visualise these points in the $(B,F)$-plane by approximating $F_n$ by $F_n \approx F_0+B_n F_1$ (an error of $O(B^2)$). This comparison is seen in figure~\ref{fig:solvability}. 
These locations where perturbation solutions are non-existent show excellent agreement with the points between adjacent branches of solutions where numerical solutions could not be calculated.

\begin{figure}
\includegraphics[scale=1]{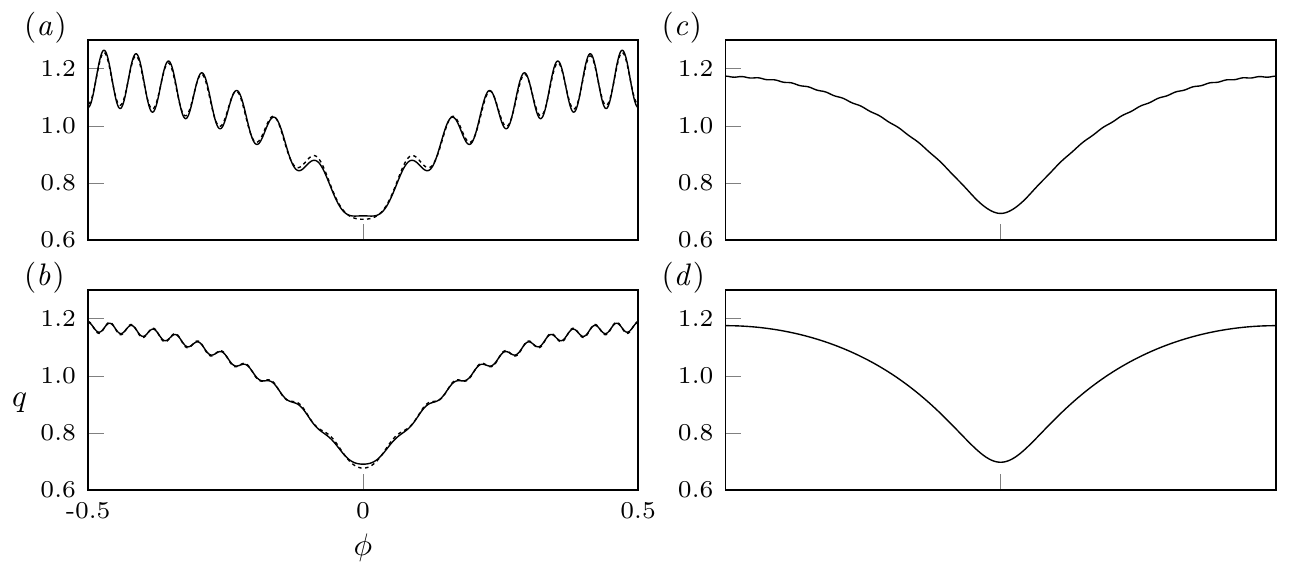}
\caption{\label{fig:ongravity} The analytical solution, $q=q_0+Bq_1+\bar{q}$, is shown (line) for the four profiles calculated in figure \ref{fig:solvability}. For comparison, numerical solutions with the same value of $B$ and $\E$ are shown dashed in insets $(a)$ and $(b)$. A value of $\lvert \Lambda_a \rvert = 0.006$ has been used for these comparisons, estimated from numerical comparisons.}
\end{figure}

Additionally, four of our analytical solution profiles, $\bar{\mathfrak{q}}$, are shown in insets $(a)$ to $(d)$ of this figure. These solutions have been selected to lie in the midpoint of the solution branch, with a Bond number of ($B=(B_n+B_{n+1})/2$). They demonstrate that the ripples obtain their greatest magnitude at the edge of the periodic domain. Note that these ripples are plotted on a zero background state. These same solutions are also shown in figure~\ref{fig:ongravity},
which includes the first two terms of the asymptotic expansion, $q_0+Bq_1$.
These have been provided to compare the magnitude of the ripples in relation to the leading-order Stokes wave.

\begin{figure}
\includegraphics[scale=1]{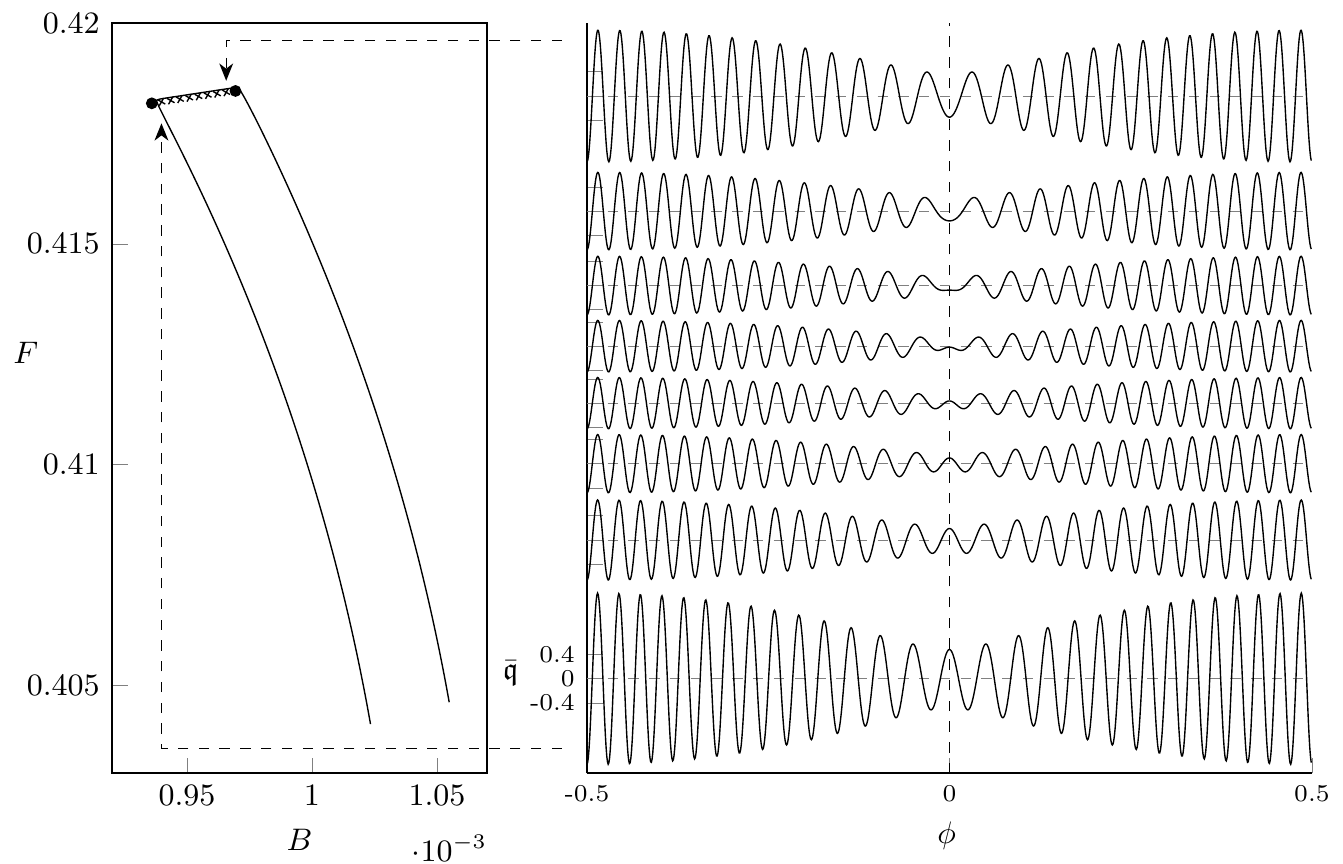}
\caption{\label{fig:onefinger} Here, for $\E=0.3804$, we plot the exponentially-small solution, $\bar{\mathfrak{q}}$, from \eqref{eq:SS12} between the two values of $B_{29}=0.0009360$ and $B_{28}=0.0009694$.
Note that the base gravity-wave is thus not shown. The eight chosen values of $B$ (crosses) are equally spaced between the values of $B_{29}$ and $B_{28}$. This corresponds to the finger $G_{28\to 29}$ found numerically by \cite{Shelton2021onthe}. A value of $\lvert \Lambda_a\rvert=1$ has been used for the constant prefactor.}
\end{figure}

In our previous numerical work, we demonstrated that as one of the solution branches was transversed, the solution develops an extra wavelength, and this was seen to occur near the top of the solution branch. We observe that the same effect occurs with our analytical solutions. This is demonstrated in figure~\ref{fig:onefinger}, in which we provide eight solution profiles equally-spaced in the Bond number between two adjacent values of $B_n$. From these, we see that as we travel from right-to left across the solution branch by decreasing the value of $B$, an additional ripple forms in the center of the domain.

\subsection{The effects of changing the energy, $\E$.}

\noindent All of the above solutions have been computed for the same fixed value of the energy, $\E=0.3804$.
We now relax this restriction by considering values of $\E$ between $0$ and $0.9$. Note that the limiting Stokes wave is not the most energetic [cf. \S{6} of \citealt{longuet-higginsTheoryAlmosthighestWave1978}] and for values of $\E$ very close to unity, there are multiple possible solutions beyond the classical Stokes wave. We shall not consider solutions too close to the highest wave ($\E>0.9$) in this work. 

\begin{figure}
\includegraphics[scale=1]{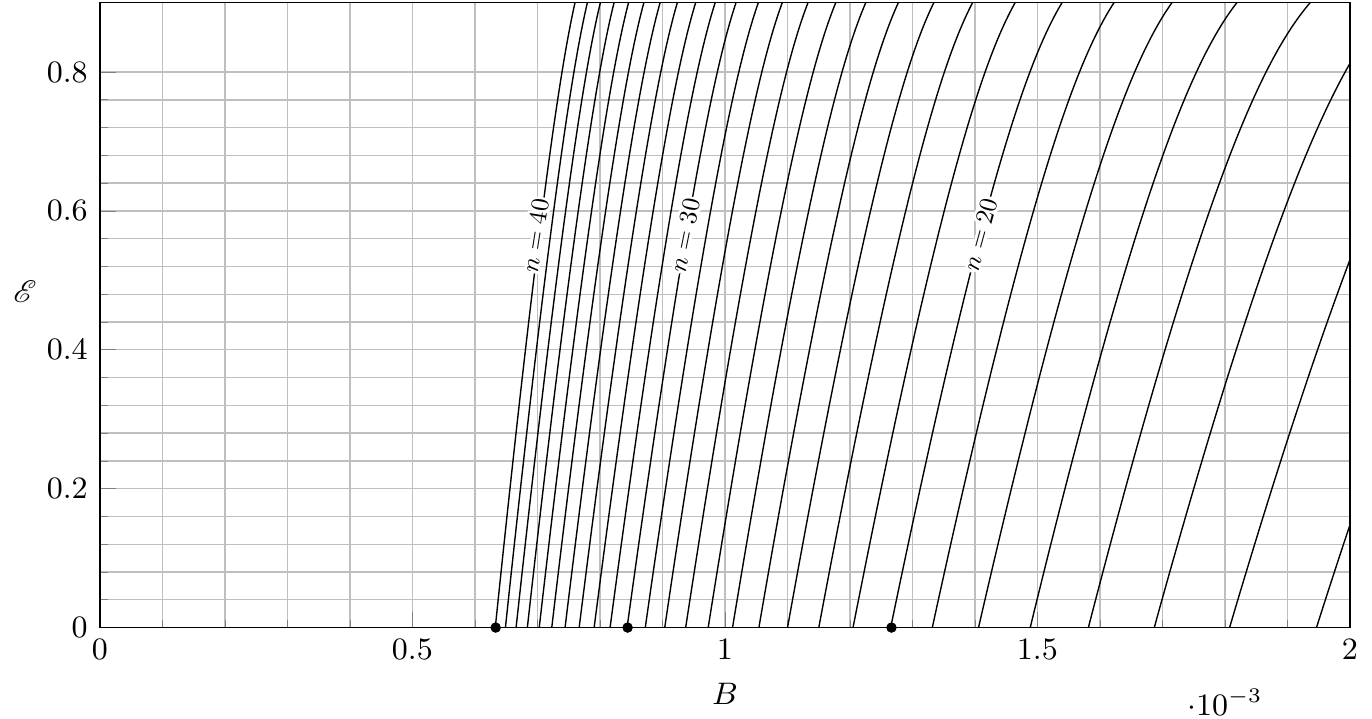}
\caption{\label{fig:Ensolv} Values for $B_n$, where the solvability condition fails, are shown for different values of the energy, $\E$. The small-$\E$ predictions by Wilton (1915) are shown by the black dots at $\E=0$ for $n=20$, $30$, and $40$.}
\end{figure}

In figure~\ref{fig:Ensolv} we show how the locations where the solvability condition fails, $B_n(\E)$, change with the energy for values of $n\leq 40$.
We note that as the energy deceases to zero and we enter the linear regime, these lines tend towards the predictions by \cite{wilton1915lxxii}. These are the discrete values of the Bond number for which two linear solutions of wave-numbers $1$ and $n$ also have the same Froude number. Thus, a single leading-order gravity-wave of the type assumed in this work is insufficient for describing Wilton's linear solutions, and is why we recover his values under this limit.

\begin{figure}
\centering
\includegraphics[scale=1]{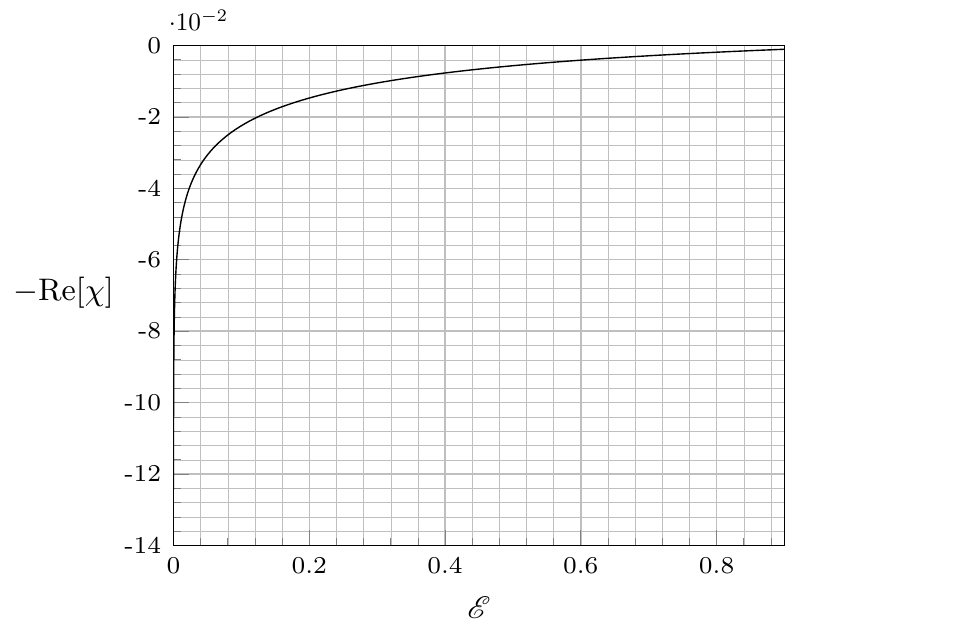}
\caption{\label{fig:Enchi} The value of $-\text{Re}[\chi]$ from equation \eqref{eq:chionphi} is shown for different values of the energy, $\E$.}
\end{figure}

We have also chosen to provide values of $\text{Re}[\chi]$ for different values of $\E$, as this controls the exponential behaviour of the magnitude of our parasitic ripples. This is shown in figure~\ref{fig:Enchi}, and shows that the constant controlling the exponential behavior of our solution increases with the energy, $\E$.

\section{Conclusions}\label{sec:Co}

\noindent We have considered the small surface-tension limit of gravity capillary waves of infinite depth.
This results in gravity-wave solutions at leading order. 
The parasitic ripples, which have a wavelength much smaller than that of the base gravity-wave, appear beyond all orders of the asymptotic expansion as their amplitude is exponentially-small in the Bond number.
The analytical solution for these from equation \eqref{eq:SS12} has been found by:

\begin{enumerate}[label=(\roman*),leftmargin=*, align = left, labelsep=\parindent, topsep=3pt, itemsep=2pt,itemindent=0pt ]
\item Observing the divergence of the Poincaré series $q=q_0+Bq_1+ \ldots$, a consequence of singularities in the analytic continuation of the leading-order solution, $q_0$.
\item Optimally truncating the divergent expansion at $N \sim 1/B$ and considering the exponentially-small remainder $\bar{q}$ by a solution of the form $q=q_0+Bq_1+ \ldots + B^N q_N + \bar{q}$.
\item Identifying the Stokes lines (which depend on $q_0$) and calculating the effect of Stokes phenomenon on the exponentially-small terms.
\end{enumerate}
We have also found a solvability condition for our problem, which fails at discrete values of the Bond number given by \eqref{eq:SS11b}.
These points were shown in figure~\ref{fig:solvability} to coincide with the discrete nature of the numerical solution branches.
Moreover, we have demonstrated that if the leading order gravity-wave is taken to be symmetric, these parasitic ripples must also exhibit symmetry about the wave crest; presenting a fundamental improvement in our understanding of the structure of these parasitic waves. 

Our results provide an analytical theory and framework for the numerical solutions detected in \cite{Shelton2021onthe}. Moreover, we have shown that, although certain details of \cite{longuet-higgins_1963_the_generation} theory of parasitic capillary ripples are correct, an exponential asymptotics approach provides verifiable asymptotic predictions, corrected functional relationships, and connection of the ripples to Stokes lines and the Stokes phenomenon.

\section{Discussion}\label{sec:Di}

\subsection{Open and resolved challenges in exponential asymptotics}

\noindent Over the past twenty years, the application of exponential asymptotics to fluid mechanical problems has been very successful in the discovery and development of new analytical methodologies \citep{boydWeaklyNonlocalSolitary1998}. However, there are a number of distinguishing features in our treatment of the parasitic ripples problem that are particularly interesting.

First, the majority of preceding works in exponential asymptotics typically rely upon the derivation of a crucial singulant function, $\chi$, for which an exact analytical form is known. In our analysis, however, the singulant in \eqref{eq:DE9} requires the complex integration of a nonlinear gravity-wave, which must be pre-computed. Moreover, the values of $\chi$ and the associated Stokes lines must be determined in the complex plane, and this has necessitated a separate study of the distribution and properties of the singularities of the Stokes wave problem \citep{crew2016new} as a precursor to the present work.

Second, there are a number of challenging steps in the exponential asymptotics analysis that we highlight here. The reader should note two interesting features. 
\begin{enumerate}[label=(\roman*),leftmargin=*, align = left, labelsep=\parindent, topsep=3pt, itemsep=2pt,itemindent=0pt ]
\item The eigenvalues, $F_n$, are divergent, but we have not had to rely upon their form in the derivation of $\mathfrak{q}_n$ in \S\ref{sec:lateorders}. 
\item Our factorial-over-power expression for $\mathfrak{q}_n$, valid only in the limit $n\to \infty$, satisfies neither the energy condition nor the periodicity conditions on $q_n$ and $q_n^{\prime}$. This is because our approximation of this divergence is only valid in the vicinity of the Stokes line about which the Stokes phenomenon occurs, rather than globally.
\end{enumerate}
Through a more detailed analysis, it is possible to derive both a factorial-over-power ansatz for $F_n$, as well as the additional terms necessary so that the late-term approximation satisfies the energetic and periodicity conditions. We provide a brief comment on the procedure, but some of these issues are more easily observed in a simpler eigenvalue problem exhibiting divergence; this will be the focus of future work by the current authors \citep{sheltondiv}.

In essence, the eigenvalue divergence produces inhomogeneous contributions to Bernoulli's equation depending on $F_n$, $F_{n-1}$, \ldots [compare \eqref{eq:OBn-noF} to \eqref{eq:OBn-bern}]. These contributions, of the form \eqref{eq:ansatzf}, will force additional components in the late-term representation of the solution. Both the periodicity and energy constraints can then be satisfied with the inclusion of further components associated with $\chi' = 0$, currently neglected following \eqref{eq:chieqn}. Once these additional divergences are included, a prediction for the eigenvalue, $F_n$, is obtained.

As it turns out however, these additional components are subdominant to the divergent ansatz \eqref{eq:ansatz} with $\chi = \chi_a(f)$ given by \eqref{eq:DE9} near the relevant Stokes lines. Consequently, these components will not influence the Stokes smoothing procedure derived in \S\ref{sec:expasymp}. We note that this is analogous to how the complex Hilbert transform, $\widehat{\mathscr{H}}[\theta_n]$, is neglected in the discussion following \eqref{eq:OBn-noF}.

\subsection{Asymmetry in steady and temporal water-waves} \label{sec:asymmetry}

\noindent It is important to note that in this work, following \cite{longuet-higgins_1963_the_generation}, we have focused on a fairly restricted view of parasitic ripples that correspond to the classical potential flow formulation of a steadily travelling wave composed of a perturbation about a symmetric nonlinear gravity wave. This assumption also follows from the class of solutions first detected by \cite{Shelton2021onthe}. 

We would expect that within this steady potential framework, it is possible to obtain general asymmetric gravity-capillary solutions exhibiting small-scales ripples in the $B \to 0$ limit. Indeed, solutions resembling this anticipated structure have been calculated by previous authors; for instance \cite{zufiria1987symmetry} considered symmetry breaking in gravity-capillary waves for moderately small values of the surface-tension coefficient. The properties of the waves in that study match those presented in this paper, as some appear to be perturbations about the asymmetric gravity waves found in \cite{zufiria1987non}. The general detection of asymmetric gravity-capillary waves remains a challenging problem (cf. \citealt{gao2017investigation}). 

However it is likely that the above relaxation of symmetry in the solutions does not lead to the typical distribution of asymmetric capillary ripples that appear on the forward-face of a steep travelling wave. In order to produce the asymmetry viewed in experimental results, it is likely necessary to consider further modifications to this theory (cf. \citealt{perlin2000capillary}). Possible extensions include accounting for the additional effects of time dependence, viscosity, or vorticity.

The problem of time-dependent parasitic waves has been studied numerically by multiple authors, such as \cite{hung2009formation}, \cite{murashige2017numerical}, and \cite{wilkening2021quasi}. For instance, \cite{hung2009formation} study a time-dependent formulation that includes vortical effects; a pure gravity wave is chosen as the initial condition and time-evolution results in the formation of parasitic ripples ahead of the wave-crest. Similar methodologies have been implemented by \emph{e.g.} \cite{deike2015capillary} in order to study the formation of time-dependent parasitic ripples in the full Navier-Stokes system using a volume-of-fluid method. We note that small-scale ripples can also occur near the crest of gravity-waves as they approach a limiting formulation, as shown by \cite{chandler1993computation} for solutions close to the steady Stokes wave of extreme form and \cite{mailybaev2019explosive} for finite depth breaking waves. In our present work, the authors are examining the application of exponential asymptotic techniques to the description of time-dependent parasitic ripples. The inclusion of time-dependence in asymptotics beyond-all-orders remains a poorly understood problem, and very few authors including \cite{chapman2005exponential}, \cite{lustri2013exponential,lustri2019three}, have considered such a complication. 

Analogously, the extension of models of gravity-capillary waves to include non-zero viscosity, vorticity, or finite depth have been considered by various authors. For instance \cite{longuet-higgins_1963_the_generation,longuet1995parasitic} and \cite{fedorov1998nonlinear} considered viscous gravity-capillary waves which exhibit asymmetry. Furthermore we would expect that a similar application of exponential asymptotics to the case of periodic finite-depth flows could be achieved; in the shallow-water limit, the results would match those presented in seminal works on generalised solitary waves in Kortewe-de Vries equations (see \emph{e.g.} \citealt{yang1996weakly,yang1997asymmetric} and chapter~10 of \citealt{boydWeaklyNonlocalSolitary1998}). It is an interesting question to consider the equivalent exponential asymptotic analysis for these more complex problems where we expect similar phenomena to arise.

\mbox{}\par
{\bf \noindent Acknowledgements}. We thank Professors Paul Milewski and John Toland (Bath) for helpful discussions, and the anonymous reviewers for their insightful comments on our work. This work was supported by the Engineering and Physical Sciences Research Council [EP/V012479/1].

\mbox{}\par
{\bf \noindent Declaration of interests}. The authors report no conflict of interest.

\appendix
\section{Singular scaling of the order $B$ quantities}\label{app:q1}

\noindent In \S\ref{sec:innerlim}, the inner limit of $q_n$ as $f \to af^*$ relied on the singular behaviour of the $O(B)$ term $q_1$.
Taking the $O(B)$ equations, we substitute $\theta_1$ from the boundary-integral equation \eqref{eq:OB-bi} into Bernoulli's equation \eqref{eq:OB-bern} to find
\begin{equation}
\label{eq:SingB1}
F_0^2  q_0^2 \frac{\de q_1}{\de f}+ \bigg[ 2F_0^2q_0q_0^{\prime} + \frac{a \i \cos{\theta_0}}{q_0}\bigg]q_1+2F_0F_1q_0^2q_0^{\prime} -a \i \widehat{\mathscr{H}}[\theta_1] \cos{\theta_0}-q_0\big(q_0 \theta_0^{\prime}\big)^{\prime}=0.
\end{equation}
The singular scaling of $\cos{\theta_0}$ can be found from equation \eqref{eq:Sing9} to be
\begin{equation}
\label{eq:SingB3}
\cos{\theta_0} \sim \frac{1}{2} \e^{a{\mathrm i}\theta_0} \sim \frac{-a{\mathrm i}F_0^2c_a^3}{4} (f-af^*)^{-\frac{1}{4}}.
\end{equation}
Thus, the term involving the complex-valued Hilbert transform $\widehat{\mathscr{H}}[\theta_1]$, which acts on the free-surface upon which $\theta_1 \sim O(1)$, is subdominant in equation \eqref{eq:SingB1}.
The same is true for the term containing $2F_0F_1q_0^2q_0^{\prime}$.
The singular scaling of the four remaining dominant terms in equation \eqref{eq:SingB1} can then be found by the results of \S\ref{sec:singqtheta}, yielding
\begin{equation}
\begin{split}
\label{eq:SingB6}
F_0^2 q_0^2 \frac{\de q_1}{\de f} &\sim F_0^2c_a^2(f-af^*)^{\frac{1}{2}}\frac{\de q_1}{\de f}, ~~~~~~~ 
 2F_0^2q_0q_0^{\prime}q_1 \sim \frac{F_0^2c_a^2}{2}(f-af^*)^{-\frac{1}{2}}q_1, \\
\frac{a{\mathrm i}\cos{\theta_0}}{q_0}q_1   &\sim \frac{F_0^2c_a^2}{4}(f-af^*)^{-\frac{1}{2}}q_1, ~~~~~~~
-q_0(q_0\theta_0^{\prime})^{\prime} \sim \frac{3a{\mathrm i}c_a^2}{16} (f-af^*)^{-\frac{3}{2}}.
\end{split}
\end{equation}
In substituting the ansatz $q_1 \sim A(f-af^*)^n$ into equation \eqref{eq:SingB1}, we then find
\begin{equation}
\label{eq:SingB7}
q_1 \sim \frac{3a{\mathrm i}}{4F_0^2}(f-af^*)^{-1}.
\end{equation}

\subsection{Inner limit of $Q_a(f)$}

\noindent To determine the value of the constant $\Lambda_a$, the analysis of which is performed in appendix \ref{sec:innermatching}, we require the inner limit of the prefactor, $Q_a(f)$, of the naive solution.
Taking $Q_a(f)$ from equation \eqref{eq:Qsolution}, we consider the singular behaviour of $q_0$ and $\e^{a \i \theta_0}$ from equations \eqref{eq:Sing8} and \eqref{eq:Sing9} to find
\begin{equation}
\label{eq:Qinnerlim}
Q_a(f) \sim \frac{-\Lambda_a a \i F_0^2 c_a^5}{2}(f-af^*)^{\tfrac{1}{4}} \exp \bigg(\int_{0}^{f} a \i  \Big[\frac{\cos{\theta_0}}{F_0^2 q_0^{3}} -F_0^2q_1-2F_0F_1 q_0 \Big]\de f\bigg) ~~ \text{as} ~~ f \to af^*.
\end{equation}
It remains to evaluate the integral in the above equation as $f \to af^*$. In considering the singular behaviour of the integrand, we find
\begin{equation}
\label{eq:Qinnerlim1}
a \i \Big[\frac{\cos{\theta_0}}{F_0^2 q_0^{3}} -F_0^2q_1-2F_0F_1 q_0 \Big] \sim  (f-af^*)^{-1}+O(1).
\end{equation}
In writing
\begin{equation}
\label{eq:Qinnerlim2}
\mathcal{P}(f)= \int_{0}^{f}a \i \Big[\frac{\cos{\theta_0}}{F_0^2 q_0^{3}} -F_0^2q_1-2F_0F_1 q_0 \Big]-(f-af^*)^{-1}\de f,
\end{equation}
and noting that $q_1 \sim \tfrac{3 a \i}{4 F_0^2}(f-af^*)^{-1}+O(1)$, we see that $\mathcal{P}(f) \sim O(1)$ as $f \to af^*$.
This formulation yields
\begin{equation}
\label{eq:Qinnerlim3}
\int_{0}^{f} a \i  \Big[\frac{\cos{\theta_0}}{F_0^2 q_0^{3}} -F_0^2q_1-2F_0F_1 q_0 \Big]\de f= \mathcal{P}(f)+\log(f-af^*)-\log(-af^*),
\end{equation}
from which we find the singular behaviour of $Q_a(f)$ to be
\begin{equation}
\label{eq:Qinnerlim4}
Q_a(f) \sim \frac{\Lambda_a  \i F_0^2 c_a^5}{2 f^*} \e^{\mathcal{P}(af^*)}(f-af^*)^{\tfrac{5}{4}} ~~ \text{as} ~~ f \to af^*.
\end{equation}

\section{An inner soution at the principal singularities}\label{sec:innermatching}

\noindent The constant $\Lambda_a$ appearing in the prefactor of $q_n$ in equation \eqref{eq:qsolution} is determined by matching the inner limit of $q_n$ with the outer limit of a solution holding near the singularity at $f=af^*$. 
In the inner region near this point, Bernoulli's equation \eqref{eq:AC-bern} holds,
\begin{equation}
\label{eq:IE1}
F^2 q^2 \frac{\de q}{\de f}+\frac{1}{2 \i}(\e^{\i \theta}-\e^{-\i \theta})-Bq\frac{\de}{\de f}\bigg(q\frac{\de \theta}{\de f}\bigg)=0.
\end{equation}

We also have the boundary-integral equation \eqref{eq:AC-bi} applying in this inner region. 
Since the complex valued Hilbert transform $\widehat{\mathscr{H}}[\theta]$ appearing in the right hand side of this operates on values of $\theta$ from the free-surface in the outer region, away from the singularity, we can use the outer expansion in powers of $B$. 
At each order in $B$, $\widehat{\mathscr{H}}[\theta_n]$ is then related to the outer solutions of $q_n$ and $\theta_n$ by evaluating the boundary-integral equation at this order. 
This gives
\begin{equation}
\begin{split}
\label{eq:IE2}
\log{(q)}+a \i \theta&=\widehat{\mathscr{H}}[\theta]\\
&=\widehat{\mathscr{H}}[\theta_0]+B\widehat{\mathscr{H}}[\theta_1]+O(B^2) \\
&=(\log{q_0}+ a\i \theta_0) + B(q_1/q_0+a \i \theta_1) + O(B^2).
\end{split}
\end{equation}
To evaluate this in the inner region, we take the inner limit of $f \to af^*$ on the right hand side. 
Exponentiating \eqref{eq:IE2} and using the scaling of $q_0$ and $\e^{a \i \theta_0}$ from \eqref{eq:Sing8} and \eqref{eq:Sing9} gives
\begin{equation}
\label{eq:IE3}
q \e^{a \i \theta}\sim -\frac{a \i F_0^2 c^4}{2} +O(B).
\end{equation}
From this, we find at leading order both $\e^{\i \theta}-\e^{- \i \theta}=2q/(\i F_0^2 c^4)-(\i F_0^2 c^4)/2q$ and $q \theta^{\prime} = a \i q^{\prime}$. Substituting these into Bernoulli's equation \eqref{eq:IE1} then gives the inner equation
\begin{equation}
\label{eq:IE4}
F^2 q^2 \frac{\de q}{\de f}-\frac{q}{F_0^2c^4}-\frac{F_0^2c^4}{4q}  -a \i Bq\frac{\de^2 q}{\de f^2}=0.
\end{equation}

\subsection{Boundary layer scalings}\label{sec:innerbl}

\noindent The width of the boundary layer at the principal upper- and lower-half plane singularities is determined by the reordering of the outer expansion $q_{\text{outer}}=q_0+Bq_1+O(B^2)$ when consecutive terms become comparable. 
Balancing $q_0 \sim Bq_1$ for simplicity, where $q_0 \sim c_a(f-af^*)^{1/4}$ from \eqref{eq:Sing8} and $q_1 \sim \tfrac{3a \i}{4F_0^2} (f-af^*)^{-1}$ from \eqref{eq:SingB7}, we find the width of the boundary layer to be $B^{4/5}$. 
Thus, we introduce the inner variable $\eta$ by
\begin{equation}
\label{eq:IE5a}
(f-af^*)=B^\frac{4}{5} \eta.
\end{equation}
Additionally, in the inner region $\bar{q}_{\text{inner}} \sim q_0$. 
By incorporating the inner variable $\eta$ with our scaling for $q_0$, we have $q_0 \sim c_a(f-af^*)^{1/4} \sim c_a B^{1/5} \eta^{1/4}$. 
This tells us how to rescale $q_{\text{outer}}$ to produce an $O(1)$ quantity, $\bar{q}_{\text{inner}}$, in the inner region, given by
\begin{equation}
\label{eq:IE5b}
q_{\text{outer}}=c_a B^{\frac{1}{5}}\eta^{\frac{1}{4}}\bar{q}_{\text{inner}}.
\end{equation}
To find the outer limit of $\bar{q}_{\text{inner}}$, we consider a series expansion as $\eta \to \infty$. 
The form of this series is determined by substituting the inner limit of the expansion for $q_{\text{outer}}$ into \eqref{eq:IE5b}, giving
\begin{equation}
\begin{split}
\label{eq:IE5c}
q_{\text{outer}}=\sum_{n=0}^{\infty}B^nq_n \sim & \sum_{n=0}^{\infty} \frac{B^n Q_a \rmGamma(n+\gamma)}{\chi_a^{n+\gamma}}\\
 \sim & \sum_{n=0}^{\infty} \frac{B^n \Lambda_a (\tfrac{\i F_0^2 c_a^5}{2f^*})\e^{\mathcal{P}(af^*)} (f-af^*)^{\frac{5}{4}} \rmGamma(n+\gamma)}{[\frac{4a \i F_0^2 c_a}{5} (f-af^*)^{5/4}]^{n+\gamma}}\\
 \sim & \sum_{n=0}^{\infty} \frac{\Lambda_a B^{\frac{1}{5}}(\tfrac{ \i F_0^2 c_a^5}{2 f^*})\e^{\mathcal{P}(af^*)}\rmGamma(n+\gamma) \eta^{\frac{1}{4}}}{(\frac{4 a \i F_0^2 c_a}{5} \eta^{5/4})^n (\tfrac{4 a \i F_0^2 c_a}{5})^{\frac{4}{5}}}.
\end{split}
\end{equation}
Here, we have used $\chi_a \sim \frac{4 a \i F_0^2 c_a}{5}(f-af^*)^{5/4}$, $\gamma = 4/5$, the singular behaviour of $Q_a$ from \eqref{eq:Qinnerlim4}, and the inner variable $\eta$ introduced in $\eqref{eq:IE5a}$. 
In denoting the constant prefactor of $\chi_a$ to be $X= 4 a \i F_0^2 c_a/5$, we find by \eqref{eq:IE5b} the expected series form for $\bar{q}_{\text{inner}}$,
\begin{equation}
\label{eq:IE5d}
\bar{q}_{\text{inner}} \sim \sum_{n=0}^{\infty} \frac{\Lambda_a (\tfrac{  \i F_0^2 c_a^4}{2 f^*})\e^{\mathcal{P}(af^*)}\rmGamma(n+\gamma)}{(X \eta^{5/4})^n X^{\frac{4}{5}}}.
\end{equation}
This suggests that in taking
\begin{equation}
\label{eq:IE5e}
z=X \eta^{5/4},
\end{equation}
the anticipated series for $\bar{q}_{\text{inner}}$ will be of the form
\begin{equation}
\label{eq:IE5f}
\bar{q}_{\text{inner}} = \sum_{n=0}^{\infty} \frac{\hat{q}_n}{z^n}.
\end{equation}

\subsection{Inner expansion}\label{sec:innerexp}

\noindent Substituting both the inner variable $\eta$ from \eqref{eq:IE5a}, and $\bar{q}_{\text{inner}}$ from equation \eqref{eq:IE5b} into the governing equation for the inner region \eqref{eq:IE4} gives
\begin{equation}
\label{eq:IE7}
c_aF_0^2\bar{q}^3\bigg(\eta \frac{\de \bar{q}}{\de \eta}+\frac{\bar{q}}{4}\bigg)-\frac{a \i \bar{q}^2}{\eta^{\frac{5}{4}}}\bigg(\eta^2 \frac{\de^2 \bar{q}}{\de \eta^2}+\frac{\eta}{2}\frac{\de \bar{q}}{\de \eta}-\frac{3 \bar{q}}{16} \bigg)=\frac{c_aF_0^2}{4}.
\end{equation}
Using the substitution $z= 4 a \i F_0^2 c_a\eta^{5/4}/5 $ presented in \eqref{eq:IE5e} results in a more convenient expansion in integer powers of $1/z$. With this, equation \eqref{eq:IE7} becomes
\begin{equation}
\label{eq:IE8}
\bar{q}^3\bigg(5z \frac{\de \bar{q}}{\de z}+\bar{q}\bigg)+\frac{\bar{q}^2}{z}\bigg(5z^2 \frac{\de^2 \bar{q}}{\de z^2}+3z\frac{\de \bar{q}}{\de z}-\frac{3 \bar{q}}{5} \bigg)=1.
\end{equation}
The outer limit of the inner solution to this equation as $z \to \infty$ is considered by the series \eqref{eq:IE5f}.
Substituting this into the inner equation \eqref{eq:IE8} yields at leading order
\begin{equation}
\label{eq:IE9}
\hat{q}_0^4=1.
\end{equation}
By considering the $O(z^{-n})$ term in \eqref{eq:IE8}, the following recurrence relation is found for $\hat{q}_n$,
\begin{equation}
\label{eq:IE10}
\begin{split}
(5n-4)\hat{q}_0^3 \hat{q}_{n}=& \sum_{k=1}^{n-1}\hat{q}_{n-k}\bigg[\hat{q}_0^2\hat{q}_k+ \sum_{p=1}^{k}\hat{q}_{k-p}\bigg(\frac{(5p-6)(5p-2)}{5}\hat{q}_{p-1}+ \sum_{j=0}^{p}(1-5j)\hat{q}_j \hat{q}_{p-j}\bigg) \bigg]\\
&+\hat{q}_0 \sum_{p=1}^{n-1}\hat{q}_{n-p}\bigg(\frac{(5p-6)(5p-2)}{5}\hat{q}_{p-1}+ \sum_{j=0}^{p}(1-5j)\hat{q}_j \hat{q}_{p-j}\bigg)\\
&+ \frac{(5n-6)(5n-2)}{5}\hat{q}_0^2 \hat{q}_{n-1}+\hat{q}_0^2 \sum_{j=1}^{n-1}(1-5j) \hat{q}_j \hat{q}_{n-j}.
\end{split}
\end{equation}

\subsection{Determining the constant $\Lambda_a$} \label{sec:innerconst}
\noindent In comparing the $n$th term of $\bar{q}_{\text{inner}}$ between representations \eqref{eq:IE5d} and \eqref{eq:IE5f}, we find the following expression for the constant $\Lambda_a$,
\begin{equation}
\label{eq:IE12}
\Lambda_a=\frac{-2 \i f^*}{F_0^2 c_a^4} \e^{-\mathcal{P}(af^*)}\Big(\frac{4 a \i F_0^2 c_a}{5}\Big)^{4/5} \lim_{n \to \infty} \frac{\hat{q}_n}{\rmGamma(n+\gamma)}.
\end{equation}
By applying Schwartz reflection principle to $q_0$, which is real-valued on the free-surface, $\text{Im}[f]=0$, we see that $c_{a=1}$ and $c_{a=-1}$ are the complex-conjugate of one another. 

The recurrence relation \eqref{eq:IE10} may then be solved numerically and yields $\lim_{n \to \infty} \frac{\hat{q}_n}{\rmGamma(n+\gamma)} \approx 1.4 \cdot 10^{-3}$. Once the secondary components of \eqref{eq:IE12} are computed, this gives a numerical value for $\Lambda_a$.


\end{document}